%% file: stack.tex
\documentclass[twocolumn]{aastex701}
\usepackage{amsmath}
\usepackage{placeins}
\usepackage{siunitx}
\input{macros}
\received{2026 August 4}
\revised{2026 October 16}
\accepted{2026 October 19}
\submitjournal{ApJ}
\shorttitle{A Composite Rest-Frame UV Spectrum of Galaxies at $5.7 < z < 6.9$}
\shortauthors{DePalma et al.}
\begin{document}

\title{A Composite Rest-Frame UV Spectrum of Galaxies at $5.7 < z < 6.9$:\\ Ionization, Outflows, and Implications for Intergalactic Metal Enrichment}
\correspondingauthor{David DePalma}
\email{depalma@space.mit.edu}
\author[0009-0003-8927-2140,gname=David,sname=DePalma]{David DePalma}
\affiliation{MIT--Kavli Institute for Astrophysics and Space Research, 77 Massachusetts Avenue, Cambridge, MA 02139, USA}
\email{depalma@space.mit.edu}
\author[0000-0003-3769-9559,gname='Robert A.',sname=Simcoe]{Robert A. Simcoe}
\affiliation{MIT--Kavli Institute for Astrophysics and Space Research, 77 Massachusetts Avenue, Cambridge, MA 02139, USA}
\email{simcoe@space.mit.edu}
\author[0000-0003-2895-6218,gname=Anna-Christina,sname=Eilers]{Anna-Christina Eilers}
\affiliation{MIT--Kavli Institute for Astrophysics and Space Research, 77 Massachusetts Avenue, Cambridge, MA 02139, USA}
\email{eilers@mit.edu}
\author[0000-0002-3120-7173,gname=Rongmon,sname=Bordoloi]{Rongmon Bordoloi} 
\affiliation{Department of Physics and Astronomy, North Carolina State University, 2401 Stinson Drive, Raleigh, NC 27695, USA}
\email{rbordol@ncsu.edu}
\author[0000-0001-9044-1747,gname=Daichi,sname=Kashino]{Daichi Kashino}
\affiliation{National Astronomical Observatory of Japan, 2-21-1 Osawa, Mitaka, Tokyo 181-8588, Japan}
\email{kashinod.astro@gmail.com}
\author[0009-0007-1062-0253,gname=Gauri,sname=Kotiwale]{Gauri Kotiwale}
\affiliation{Institute of Science and Technology Austria (ISTA), Am Campus 1, 3400 Klosterneuburg, Austria}
\email{Gauri.Kotiwale@ist.ac.au}
\author[0000-0003-2895-6218,gname=Jorryt,sname=Matthee]{Jorryt Matthee}
\affiliation{Institute of Science and Technology Austria (ISTA), Am Campus 1, 3400 Klosterneuburg, Austria}
\email{jorryt.matthee@ist.ac.au}
\begin{abstract}
We present a stacked JWST/NIRSpec spectrum of 104 galaxies at $\lambda_{\rm rest}=1200$--$2000$\ \AA\ to study nebular emission and absorption from the interstellar medium (ISM) and stellar winds. The sample is drawn from known \oiiiforbid\ emitters with mean redshift $z=6.50$, absolute UV magnitude $M_\mathrm{UV}=-20.7$, and stellar mass $M_{\star}=10^{8.93}\;\mathrm{M_\sun}$. The stacked continuum closely resembles lower-redshift Lyman Break Galaxy composites, with a few notable differences. Ly$\alpha$ has comparable strength and is redshifted relative to the \oiiiforbid\ systemic frame, consistent with production in gaseous outflows. High-ionization emission lines including \oiiisemi, \ciiisemi, and \heii are slightly stronger than in lower-redshift stacks but exhibit narrow widths, disfavoring ionization from a broad-line AGN. Emission-line ratios slightly favor ionization by a hard stellar continuum, though a contribution from a narrow-line nuclear source cannot be strictly excluded. Numerous strong absorption lines at both high and low ionization are blueshifted by $\sim200$\ \kms, again indicating  outflows driven by stellar winds or entrained metal-enriched ISM. A simple model suggests that at $z\sim6.5$ most star-forming galaxies are too young for winds at these velocities to propagate beyond $\sim3$--$9$ proper kpc, implying that the observed \oiiiforbid emitter population alone cannot explain the frequency of heavy-element absorbers seen in quasar spectra. In this picture, early enrichment of the intergalactic medium likely arises from undetected low-mass galaxies ($M_\star\sim10^5\text{--}10^6\;\mathrm{M_\sun}$) clustered around the more massive systems detected in JWST/NIRCam WFSS surveys.
\end{abstract}

\section{Introduction}

The rest-frame UV spectra of galaxies ($\lambda\approx1000$--$2000$\,\AA) contain a rich set of strong emission and absorption features that serve as diagnostics of interstellar heavy-element abundances, ionization conditions, gas dynamics, and feedback. Early and more recent studies of Lyman Break Galaxies at $z\sim2-3$ demonstrated the utility of stacking low-SNR spectra with $R\sim1000$ ($\Delta v\sim300$\,km\,s$^{-1}$) into high-SNR composites \citep{shapley, 2018ApJ...853...87R, aurora_stack, 2020MNRAS.493.3194P}. One of the most striking features of such composites is the systematic blueshift of interstellar absorption lines and the corresponding redshift of Ly$\alpha$ relative to the systemic rest frame. The absorption features in particular have been interpreted as signatures of galactic outflows, whose absorbing column must cover a substantial fraction of the stellar disk to generate large rest-frame equivalent widths $W_r$.

With the advent of \textit{JWST}, thousands of galaxies have now been observed within the reionization epoch at $z=6\text{--}14$, and several groups have constructed composite spectra to study topics ranging from ionized outflows in emission \citep{JADES_Outflows_2024}, to AGN signatures \citep{JADES_LowMassBH}, to ISM chemical enrichment \citep{JADES_ChemEnrich_2025, Hayes2025OnPhotons, Roberts-Borsani2026JWSTPathway}, to trends with Ly$\alpha$ emission \citep{JADES_LyA_nonLyA}, and general ISM properties \citep{CEERS_ISM, JWST_Spectra_z9_14}. These initial stacks do not have sufficiently high SNR in the rest-UV continuum to measure interstellar absorption-line equivalent widths and velocities, nor the spectral resolution required to determine accurate offsets relative to systemic redshifts. Initial studies of absorption lines achieved high SNR either by focusing on a single transition across cosmic time \citep[\ion{Mg}{2} from $1\lesssim z\lesssim10$;][]{mgii_absorption_outflows} or selecting individual UV-bright galaxies that independently exhibit high SNR \citep[][]{Zhu2026EarlyReionization}. More recently, \citet{2025ApJ...992..191G} constructed a composite spectrum at $z=6.0-9.4$  from galaxies with secure redshifts and G140M coverage in the JADES \citep{2026ApJS..283....6E}, CEERS \citep{2025ApJ...983L...4F}, and AURORA \citep{2025ApJ...980..242S} surveys.  These early papers span a wide range in redshift and continuum luminosity. They have uncovered evidence of early metal enrichment in the ISM, as well as outflows in low-ionization lines. However, \citet{2025ApJ...992..191G} report evidence that low-ionization interstellar absorption lines, which trace outflows, show reduced equivalent width and velocity offset from the rest frame compared to similarly-constructed low redshift composites.

In contrast, the MASQUERADE survey \citep{masquerade} obtained high-SNR NIRSpec observations with the G140M grating for a sample of $>100$ galaxies at $5.7<z<6.9$, for which systemic redshifts had already been established from rest-frame optical ISM lines including \oiiiforbid\ and H$\beta$ based on NIRCam wide-field slitless (WFSS) observations \citep{kashino_eiger}. The MSA integrations were designed to meet strict continuum-SNR and spectral-resolution requirements for individual objects, enabling robust detection of intergalactic medium transmission blueward of Ly$\alpha$ in regions ionized by foreground quasars \citep{masquerade, yue_masquerade}. These same qualities make MASQUERADE spectra especially suitable for stacking analyses, with the proviso that its sample galaxies are on average more luminous than composites selected from other surveys.  As shown below, our stacked mean galaxy has $M_{\rm UV}=-20.7$, one magnitude brighter than that of \citet{2025ApJ...992..191G} who report $M_{\rm UV}=-19.7$ \citep[where $M^*_{\rm UV}=-21.15$;][]{Bouwens2021NewEfficiency}.

Motivated by the routine detection of intervening heavy-element absorption in $z>6.5$ quasar spectra \citep{davies_xqr30, becker_oi, christiansen_metals, chen_z6_mgii}, the frequent signatures of galactic outflows in $2<z<4$ LBGs \citep{shapley, Steidel2018_CGMComposite, Steidel2010_Outflows_z2to3, Erb2012_OutflowScaling, Du2018_LyAComposite}, and the emergence of new galaxy populations at $z>6$ with comparatively low stellar and inferred halo masses \citep{Matthee2023EIGER.JWST}, it is natural to consider whether winds from these galaxies can escape into the intergalactic medium (IGM) and drive early intergalactic enrichment. However, such speculation must be balanced against the limited cosmic time available for winds to traverse large distances and the even younger inferred star-formation ages of these systems.

A stacked spectrum of the galaxies observed in the MASQUERADE survey offers an opportunity to test this hypothesis by measuring the strengths and velocity offsets of interstellar absorption features associated with galactic winds. Combined with demographic information from NIRCam rest-frame optical photometry and spectroscopy, we can characterize outflows in the epoch of reionization (EoR) and assess the extent to which the observed galaxy population contributes to cosmic metal enrichment inferred from quasar absorption systems.

Throughout this work we adopt the Planck 2018 cosmological parameters \citep{Aghanim2020PlanckParameters}: $H_0 = 67.4 \pm 0.5\;\mathrm{km\,s^{-1}\,Mpc^{-1}}$ and $\Omega_{0,\mathrm{m}}=0.315\pm0.007$.

\section{Methods}

\begin{figure*}
    \hspace*{-0.5cm}
    \includegraphics[width=\textwidth]{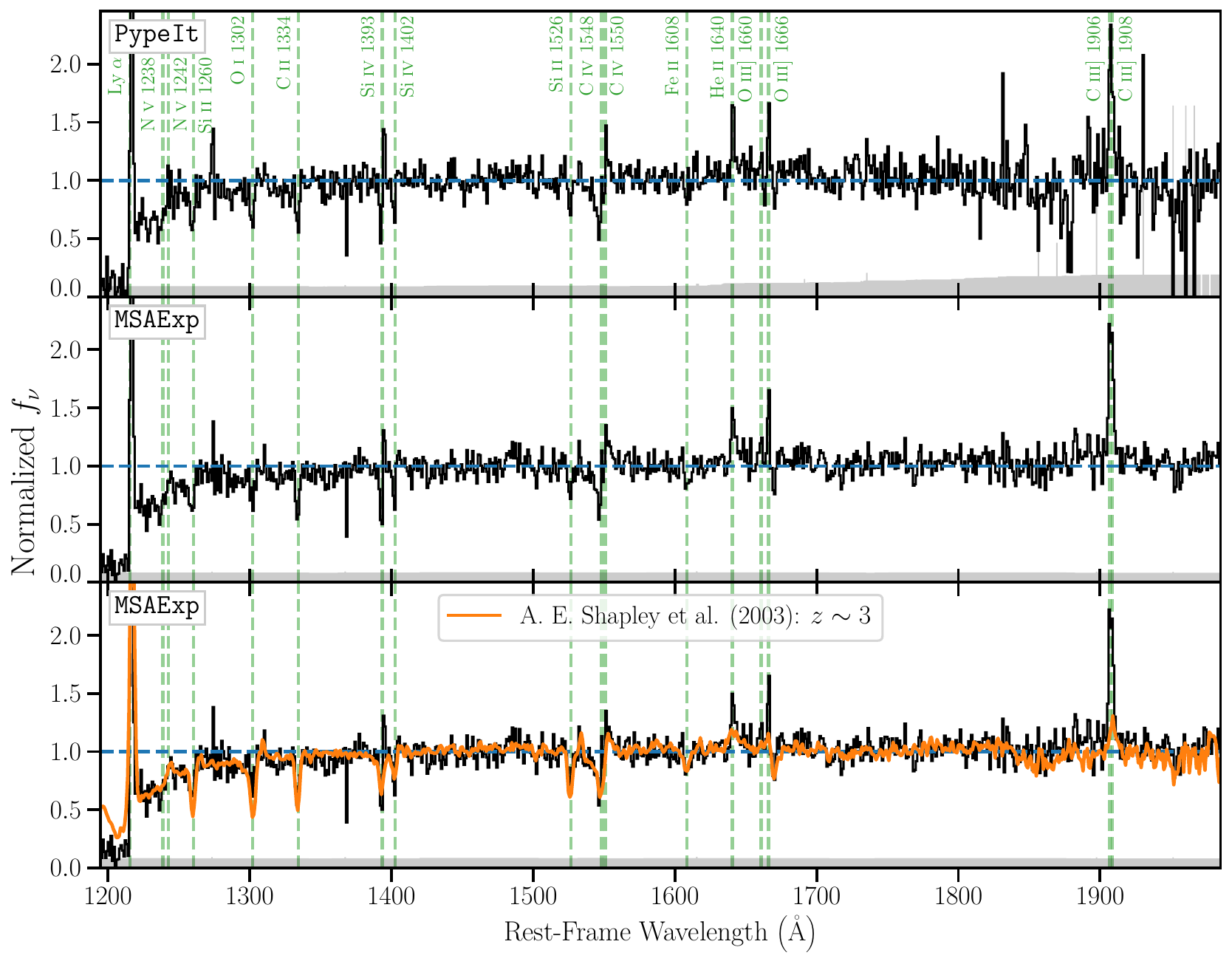}
    \caption{Stack of 104 MASQUERADE spectra as reduced by the \texttt{PypeIt} (top) and \texttt{MSAExp} (middle) pipelines. The bottom panel compares our {\tt MSAExp} reduction with the $z=3$ stack of Lyman-break galaxies from \citet{shapley}, illustrating the similarity of Ly\;$\alpha$ and also interstellar absorption lines as labeled in green. The semiforbidden \ciiisemi and \oiiisemi lines and \heii $\lambda$1640\AA\ are stronger in the high-redshift stack. A higher noise level and several apparent emission features are visible in the {\tt PypeIt} stack near to / redward of $\approx 1850$\AA\ that are not seen in the {\tt MSAExp} stack. These are spurious, and caused by edge effects from truncation of {\tt PypeIt} spectra in second-order G140M overlap regions, which reduces the number of contributing galaxies.}
\label{fig:stairsreductions}
\end{figure*}

\subsection{Spectroscopy}\label{data}

The input data for our stacks originate from the MASQUERADE survey \citep[][PID~4713]{masquerade}, which observed 104 galaxies at $5.7<z<6.9$ within the quasar fields J0100+2802 and J1148+5251 with JWST/NIRSpec’s G140M grating plus F070LP blocking filter in the MSA configuration. The targets are drawn from two of the six fields observed by the EIGER survey \citep{EIGER_I}, selected because of galaxy overdensities identified behind the survey quasars. EIGER observed these fields with NIRCam in WFSS mode and the F356W blocking filter, identifying high-redshift objects based on \oiiiforbid 4960, 5008 and H$\beta$ emission. While the sample was selected on emission properties, we fitted {\tt Prospector} models to three-band galaxy photometry and line fluxes (described in more detail below) to understand its demographics. The sample has a mean redshift and standard deviation of $\langle z\rangle\approx6.3169\pm0.3864$, UV magnitude $\langle M_\mathrm{UV}\rangle\approx-19.45\pm1.19$, and stellar mass $\langle\log{M_\star/\mathrm{M_\odot}}\rangle\approx8.33\pm0.76$. The distribution of SFR is less symmetric; its median, sixteenth and eighty-fourth percentile are $\mathrm{SFR_{H\;\beta}}\approx{3.4}_{-2.5}^{+9.0}\;\mathrm{M_\sun\,{yr}^{-1}}$.

MASQUERADE's followup slit mask design prioritized brighter targets from the EIGER fields (F115W $\lesssim\qty{27}{mag}$) with known spectroscopic redshifts \citep{kashino_eiger} located in the background or immediate environment of the luminous target QSOs. Consistent with the survey’s primary science goals, each mask was observed for 7.7\,hr to detect continuum flux blueward of Ly$\alpha$, yielding unusually high-SNR individual spectra for the G140M grating at $R\sim1000$ (median $\sim0.85$ and ranging from $<0.05$ to $\approx 4$ per resolution element in the continuum) and covering the rest-UV range $\lambda\approx1200$--$2000$\,\AA. 

Further details on the data reduction procedures are provided in the MASQUERADE survey paper. In this work we employ two independent reductions of the raw frames into 1D spectra to evaluate consistency and assess the possibility of systematic effects introduced by the pipelines. The first reduction used a customized version of the \texttt{PypeIt} package \citep{Prochaska2020PypeIt:Pipeline}, while the second employed an automated implementation of the \texttt{MSAExp} pipeline \citep[v0.3][]{msaexp, msaexp_science, msaexp_heintz} as incorporated into the Dawn JWST Archive. When quoting statistics from each different reduction below, we list values from the \texttt{MSAExp} reduction directly in the body, and corresponding \texttt{PypeIt} values in parentheses.

\subsection{Stacking}\label{Stacking}

To prepare the individual spectra for stacking, we shifted each galaxy into its rest frame using the [O\,\textsc{iii}] redshifts measured in NIRCam WFSS spectra from the EIGER survey \citep{Matthee2023EIGER.JWST, kashino_eiger}. Each spectrum and its associated error vector was then normalized by the median flux (inclusive of spectral lines) at $\lambda_{\rm rest}>1300$\,\AA.

The error vectors produced by both pipelines appear to underestimate the empirically measured noise level in our spectra. We therefore rescaled the quoted errors over the wavelengths of interest by calculating each spectrum's median absolute deviation and converting this into a standard deviation under the assumption that the errors are approximately normally distributed. This single rescaled median error (one measurement per spectrum) is used as the optimal weighting factor for that object's contribution to the composite.

Next, we rebinned all spectra onto a common rest-frame wavelength grid with $1$\,\AA\ pixels, which corresponds to $\sim250$\,km\,s$^{-1}$ at the blue end of our coverage and $\sim150$\,km\,s$^{-1}$ at the red end. Finally, the prepared spectra were stacked by computing an inverse-variance-weighted mean of all raw pixels from the 104 galaxies in our sample that fall within each binned pixel (which differs slightly between {\tt PypeIt} and {\tt MSAExp}). We estimate the error on each stacked pixel as:
\begin{equation}
    \sigma_{\rm stacked\ pix} = \sqrt{\sum_{\rm galaxy\ pix} (1/\sigma^2_{\rm galaxy\ pix})}
\end{equation}

Figure \ref{fig:stairsreductions} illustrates the two stacked spectra (one for each reduction). The weighted mean absolute UV magnitude of the stacked galaxy spectrum is $M_\mathrm{UV}=-20.7$ for the {\tt MSAExp} reduction (and $M_\mathrm{UV}=-20.8$ for {\tt PypeIt}), and the weighted mean redshift of the stacked galaxy spectrum is $z_{\rm stack}=6.50$ for {\tt MSAExp} ($z_{\rm stack}=6.56$ for {\tt PypeIt}). The very slight difference in mean absolute UV magnitude or mean redshift results from small disparities in the per-pixel SNR (and hence stacking weights) for the two extraction codes as well as pipeline choices about the spectral wavelength bounds. We measure the continuum signal-to-noise ratio in the stacks red of \qty{1300}{\text{\AA}} as 10.25 (7.83).

\input{equivalentwidthdispersion}

Rest wavelengths of several strong interstellar absorption and emission lines are indicated in green. Although many individual spectra have low SNR, the {\tt PypeIt} and {\tt MSAExp} composites display consistent continuum shapes and similar emission and absorption features, indicating that our results are relatively insensitive to systematic effects introduced by pipeline extraction, wavelength calibration, or flux calibration.

We caution that, for galaxies at the high end of our redshift range, the red portion of the stacked spectrum ($\lambda_{\rm rest}\gtrsim1800$\,\AA) overlaps the second-order footprint of the same object's G140M spectrum, making flux measurements in this region potentially susceptible to contamination from blue light in the second order. {\tt PypeIt} discards the contaminated spectral regions; {\tt MSAExp} extracts these wavelengths but does not attempt to correct the extracted flux for second-order contamination.  The effect should be modest because of Gunn--Peterson absorption blueward of Ly$\alpha$, which suppresses object flux at the blue end of the second order. The \ciiisemi\ line, which falls in this part of the spectrum, remains useful as a consistency check on the systemic redshift of the stack (i.e., to confirm that there are no systematic redshift offsets between the NIRCam [O\,\textsc{iii}] redshifts and \ciiisemi\ from NIRSpec; see Section \ref{sec:emissionvelocity}). Measurements of \ciiisemi\ fluxes or equivalent widths should, however, be interpreted with caution. 

An additional related phenomenon causes a drift in redshift with increasing wavelength at the red extreme of the composite, because each raw spectrum spans a different range in rest-frame wavelength depending on its redshift and slit position. We do not analyze the stack at rest wavelengths bluer than $1200$\AA, ensuring that every galaxy contributes to each pixel on the left of Figure \ref{fig:stairsreductions}, thereby minimizing this effect. On the red side, some objects terminate at shorter wavelengths than others, because of detector boundaries and chip gaps as well as {\tt PypeIt}'s conservative trim of second-order regions. Because {\tt MSAExp} does not perform this trim, the effect is minimal with variations in per-pixel redshift of only $\Delta z\sim \pm 0.02$ from $\lambda_{\rm rest}=1200$ to 1950\AA. For {\tt PypeIt}, the effect is more pronounced as fewer galaxies contribute to the reddest pixels, from a mean of $z=6.4$ at $\lambda_{\rm rest}=1600$\AA\ to $z=6.6$ at $\lambda_{\rm rest}=1950$\AA.  This is also seen in the reduced SNR of {\tt PypeIt}'s stack at red wavelengths.

\begin{figure*}
    \centering
    \includegraphics[width=0.7\textwidth]{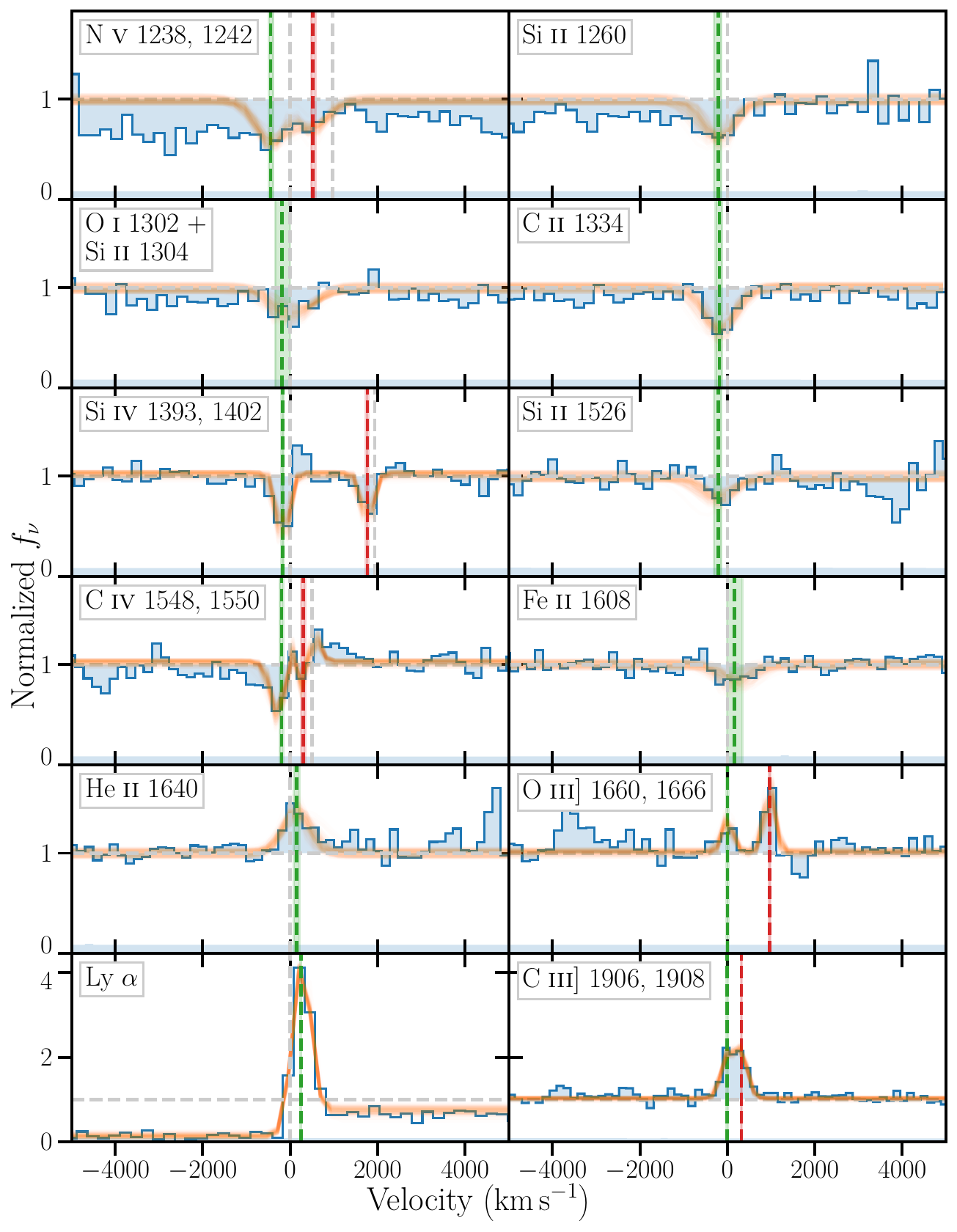}
    \caption{Velocity stack of all ions listed in Table \ref{tab:linemeasurements}, with model fit profiles drawn at random from respective posterior distributions. Vertical gray dashed lines indicate the zero-velocity rest frame. Green and red dashed lines indicate the best-fit centroids for each ion or transition, with uncertainties shown by shaded regions of the same color. Singlet transitions are shown with one green line, and doublets are shown in green and red for the two respective components. Here we show the {\tt MSAExp}-reduced composite spectrum; an equivalent plot showing the {\tt PypeIt} reduction is presented in the Appendix.}
    \label{fig:plottransitions}
\end{figure*}

\subsection{Comparison to \texorpdfstring{$z=3$}{z = 3} Lyman Break Composite}

In the bottom panel of Figure \ref{fig:stairsreductions}, we over-plot the $z\sim3$ stack of \citet{shapley} for reference. \citet{shapley} stacked 811 Ly-break galaxies with $\mathcal{R}\sim24.6$ and dust-corrected $\mathrm{SFR}\sim\numrange[range-phrase=\text{--}]{30}{50}\;h^{-2}\mathrm{M_\sun\,{yr}^{-1}}$. At $z\sim3$, $\mathcal{R}\sim24.6$ corresponds to $M_{1600}\sim-22.5$. Although this paper does not report its sample's stellar mass, in \citet{Shapley2001TheGalaxies} the same team fit SEDs to 74 bright galaxies in the associated parent sample and found a median $\log{\left(M_\star/\mathrm{M_\sun}\right)}\approx10.1$. Their sample is therefore slightly brighter / more massive  than MASQUERADE on average, but comparable in number and SNR, and form a reasonable standard for comparison.

The two stacks are remarkably similar, particularly in their interstellar absorption features: the same set of ions is present, and we will discuss below that both stacks show a comparable blueshift of most absorption features relative to the systemic frame. Both also show continuum absorption redward of Ly$\alpha$ at $\lambda\approx1220$--$1250$\,\AA, which may arise from optically thick H\,\textsc{i} in the same outflowing gas producing the heavy-element absorption \citep{msaexp_heintz}.

In high-redshift quasar spectra, similar damping wings have been used to measure the neutral fraction of the IGM \citep[e.g.,][]{banados_pisco}. However, in the $z\sim 3$ composite, such absorption is unlikely to reflect a neutral surrounding IGM because reionization has long been completed \citep{Keating2020LongHydrogen, Bosman2022HydrogenSample}. Even $z\sim 6.5$ is lower than the redshift at which most quasars begin to show evidence of strong IGM damping wings \citep{annreviews}, though it is possible that the higher-redshift contributors to our stack approaching $z=6.9$ could be affected. Measuring the late-stage chronology of reionization through galaxies' Ly-$\alpha$ damping wings would be challenging because the damping wings would be indistinguishable from intrinsic absorption. (though \citet{Davies2025PossibleAbsorption} examine how to distinguish IGM absorption from a foreground DLA, and \citet{Meyer2025ProbingFields} calculate a galaxy-based constraint on the IGM's opacity as reionization ended.) Still, the strong similarity between the $z=3$ and $z=6.5$ stacks suggests that {\em on average} there is little evidence for enhanced neutral IGM absorption in proximity to MASQUERADE's galaxies.

It is apparent from Figure \ref{fig:stairsreductions} that UV emission lines are stronger in the high-redshift MASQUERADE stack, including \heii\ $\lambda1640$, \oiiisemi\ $\lambda\lambda1660,1666$, and \ciiisemi\ $\lambda1909$ (though, as noted above, the latter may be affected by second-order contamination). Visual comparisons of the spectra could be influenced by differences in the effective spectral resolution of these stacks. \citet{shapley} estimate their LBG composite resolution at $\sim400\text{--}700$ \kms\ FWHM, whereas our NIRSpec composite ranges from $R\sim600$ near \lya to almost $R\sim1000$ near \ciiisemi\, which could cause emission lines to appear taller and more narrow. In subsequent discussions about emission line properties, we focus on integrated line fluxes and equivalent widths to minimize the impact of uncertain stack resolution on our conclusions.

\subsection{Emission and Absorption Line Fits}

The composite spectra in Figure \ref{fig:stairsreductions} show clear signatures of emission and absorption from multiple species listed in Table \ref{tab:linemeasurements}. To parameterize these features, we fit model profiles to each transition using Markov chain Monte Carlo (MCMC) walkers implemented in \texttt{emcee} \citep{Foreman-Mackey2013EmceeHammer} to sample the posterior distributions of the model parameters.

Singlet absorption transitions (e.g., \cii\ $\lambda1334$ or \oi\ $\lambda1302$) were represented by a Gaussian profile with fit parameters of velocity centroid ($v$; with zero point defined by \oiiiforbid 4960, 5008), amplitude ($A$), width ($\sigma$), and a floating constant continuum offset ($a$), all with uniform priors ($v \in [-500,500]$ \kms; $A \in [0,1]$; $\sigma \in [0,500]$ \kms; $a \in [0.95,1.05]$; four free parameters). Doublet or multiplet transitions (e.g., \siiv\ $\lambda\lambda1393,1402$ or \siii\ $\lambda\lambda1260,1526$) were fit using multiple Gaussians sharing a common velocity centroid $v$ and width $\sigma$, but with independent amplitudes ($A_1,A_2$) and continuum offsets ($a_1, a_2$) (six free parameters). The only exception is Ly$\alpha$, which is a single line but uses two continuum parameters $a_1$ and $a_2$ to account for the jump across its spectral break.

Doublets possess a prior constraint on the ratio between their components' amplitudes $A_1/A_2$. In absorption, the minimum is the ratio of the components' oscillator strengths. In emission, the minimum is the ratio of the components' weighted transition probabilities, $g_kA_{ki}$. The ratio's parameter is free between the unsaturated minimum and unity as the saturated maximum.

The \oi\ $\lambda1302$ transition is blended with adjacent \siii\ $\lambda1304$ absorption and thus required special treatment. To separate the two components, we modeled the blended \siii\ by fixing its $v$ and $\sigma$ using constraints from the earlier fit to \siii's $1260$\,\AA\ and $1526$\,\AA\ components and allowed its amplitude $A$ to vary between prior bounds set by the oscillator-strength ratio (unsaturated case) and unity (fully saturated case). The \oi\ absorption was then modeled jointly with this range of allowed \siii\ amplitudes to produce the posteriors quoted in Table \ref{tab:linemeasurements}.

For \ion{C}{4} 1548, 1550, we applied a P-Cygni profile. We fit two doublets: one in absorption with free velocity and width plus another in emission fixed at stationary velocity and with our stack's estimated resolution as the width. Each doublet can freely vary the ratio of its components' amplitudes between the unsaturated minimum and saturated unity maximum. In Table \ref{tab:linemeasurements}, we list the \ion{C}{4} absorption and emission doublets under their respective headings even though we fit both components together in a single run.

\begin{figure*}
    \centering
    \includegraphics[width=0.65\textwidth]{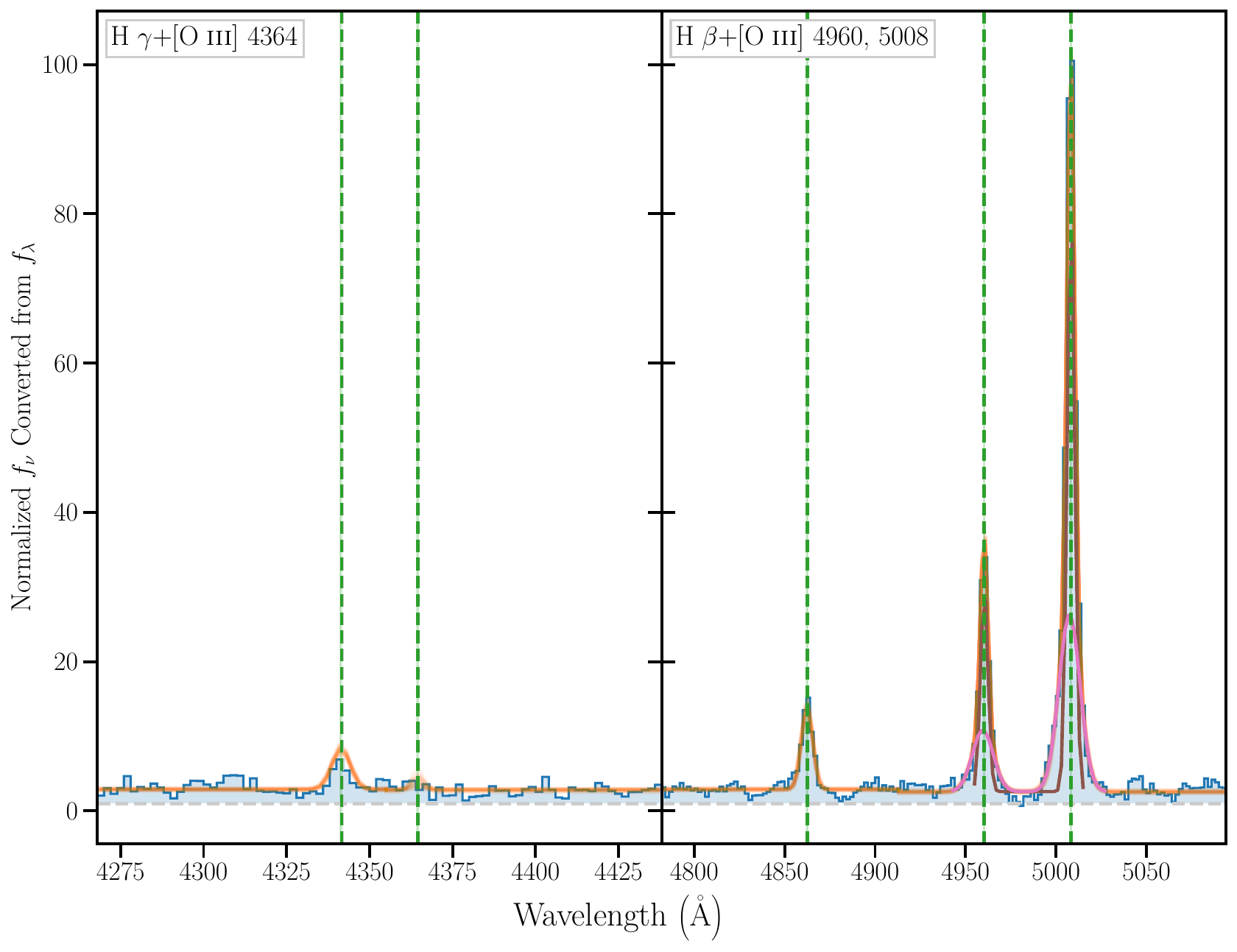}
    \caption{Stack of JWST/NIRCam WFSS rest-frame optical H$\gamma$ and \oiiiforbid $\lambda 4364$ lines (left) and the \oiiiforbid $\lambda\lambda 4960,5008$\ \AA\ doublet and H$\beta$ lines (right) for the MASQUERADE sample. Objects and weights used to generate this stack are identical to what was used for the NIRSpec stacks presented in Figure \ref{fig:stairsreductions}. The H$\beta$/H$\gamma$ ratio indicates low dust extinction ($A_V\approx{0.50}_{-0.50}^{+1.10}$;  \citet{Calzetti2000TheGalaxies}), and the non-detection of 4364\ \AA\ emission from auroral \oiiiforbid implies a low electron temperature characteristic of stellar radiation fields (rather than AGN). Here, we show the composite spectrum with weights from \texttt{MSAExp}; we provide an equivalent plot with weights from \texttt{PypeIt} in the Appendix.}
    \label{fig:NIRCam_masquerade_stack}
\end{figure*}

We maximized the log-likelihood by minimizing a merit function closely resembling the $\chi^2$ statistic. The only distinction from a true $\chi^2$ is an additional term that allows for rescaling of the per-pixel error vector \citep{Foreman-Mackey2013EmceeHammer}. We define fitting windows of $\pm500$\,km\,s$^{-1}$ for singlet transitions (or for each line of a widely spaced multiplet); for closely spaced doublets the bounds extend from $-500$\,km\,s$^{-1}$ to the blue of the blue component to $+500$\,km\,s$^{-1}$ to the red of the red component.  Figure \ref{fig:plottransitions} presents velocity plots of each transition with draws from its fit posteriors.

\begin{figure*}
    \plottwo{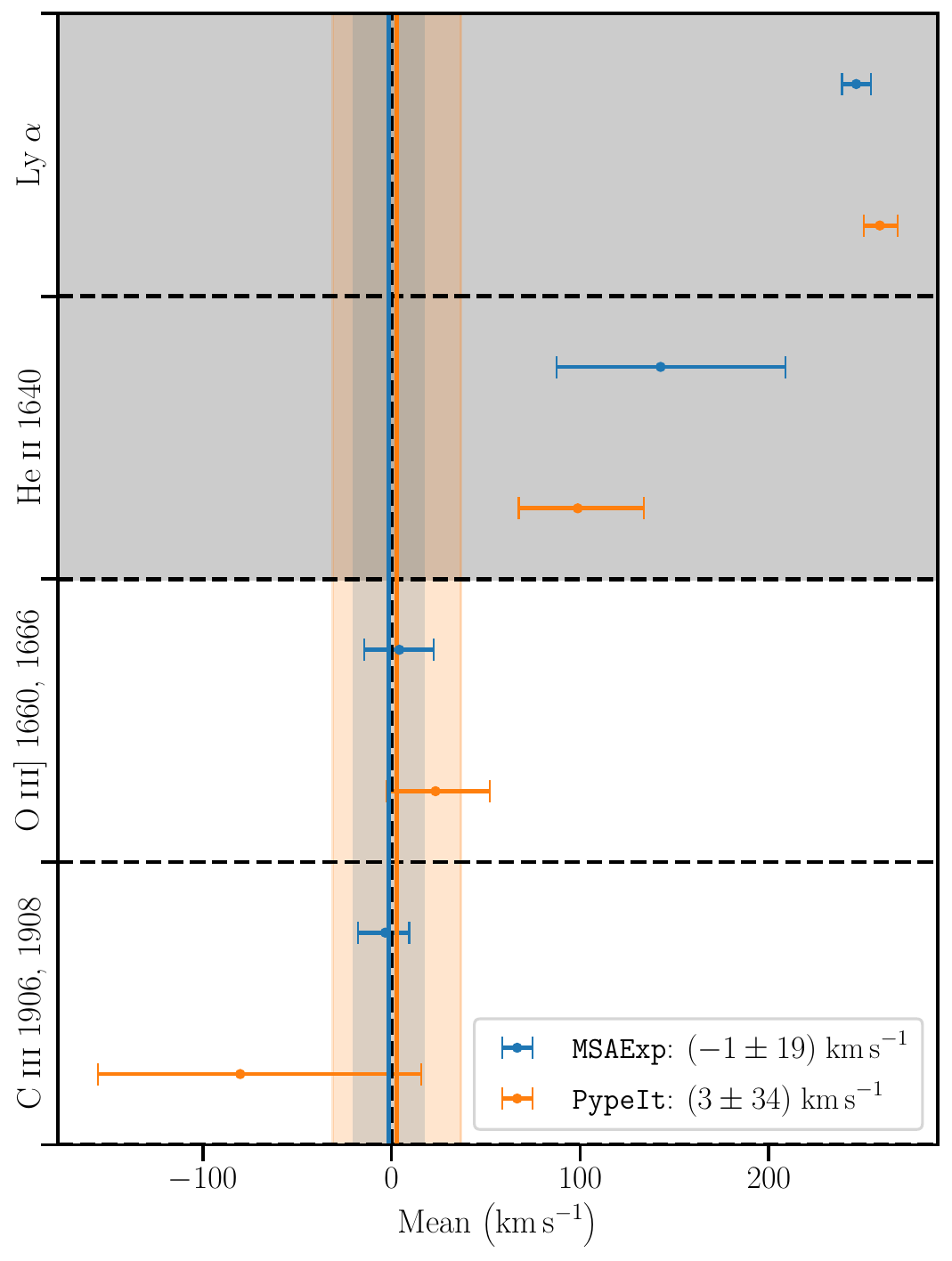}{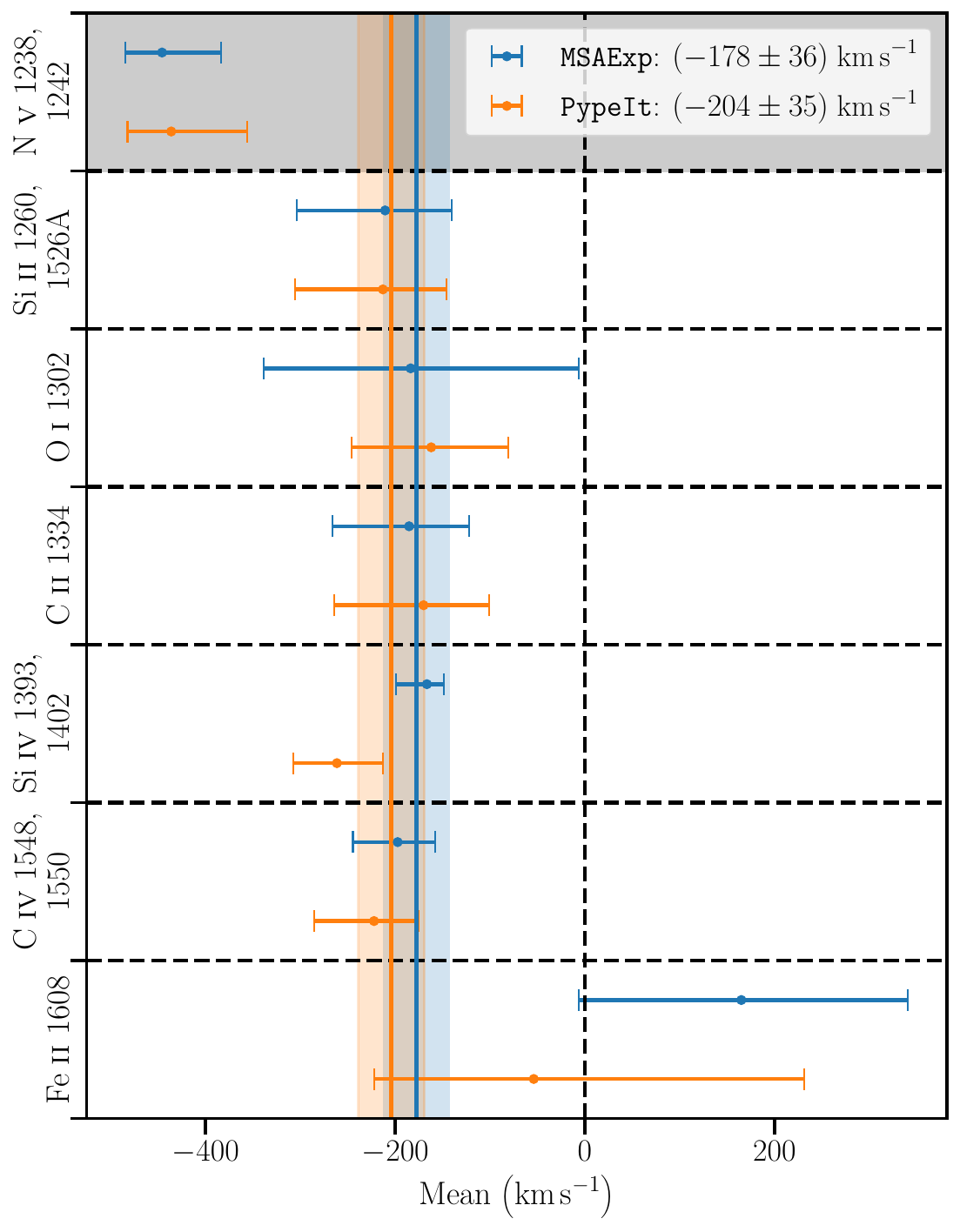}
    \caption{(\textit{Left}) Summary plot showing individual emission line centroids and the velocity uncertainty on the centroid measurement. We use \oiiisemi and \ciiisemi, which should closely track the ISM systemic rest frame, to cross-validate the wavelength solutions for our NIRSpec (MASQUERADE) and NIRCam (EIGER) data reductions. These lines are consistent with zero velocity offset relative to the rest frame, which we define by convention using the \oiiiforbid doublet wavelength. The vertical solid shaded regions show the inverse-variance weighted average of these line centroids. \lya and \heii are redshifted relative to the rest frame. In the case of \lya, this is interpreted as back-scattered emission from a receding shell behind the galaxy. \heii does not share this behavior but may be affected by foreground resonant absorption. We therefore exclude Ly\;$\alpha$ and \ion{He}{2} from the average velocity and indicate so by shading them gray. (\textit{Right}) This plot shows a summary of fit centroids for absorption lines as labeled.  The vertical colored lines and regions show respectively the mean and standard error on the mean for all absorption lines, which is blueshifted by $180\text{--}200$ \kms relative to the rest frame. \nv exhibits a larger blueshift but we view it as unreliable because of continuum absorption in the same spectral region. It is therefore shaded gray and not included in the average outflow velocity calculation.}
    \label{fig:emline_offsets}
\end{figure*}

\section{Results}

\subsection{Emission Line Velocity Structure\label{sec:emissionvelocity}}

Because individual spectra were shifted into the rest frame using tabulated [O\,\textsc{iii}] redshifts from NIRCam WFSS spectra (redshift uncertainty $\Delta z\sim100$\,km\,s$^{-1}$; \citealt{kashino_eiger}), it is useful to measure velocity offsets of UV emission lines that should coincide with the systemic redshift as a consistency check. Any offset from zero could indicate systematic differences between instruments or reductions, and would complicate interpretation of absorption-line velocity offsets.

The two most reliable systemic indicators are the semiforbidden \ciiisemi\ $\lambda1909$ and \oiiisemi\ $\lambda\lambda1660,1666$ transitions. As shown in Table \ref{tab:linemeasurements} and Figure \ref{fig:emline_offsets}, these fits are consistent with zero velocity offset. Averaging the fits for both lines, we find $\Delta v=\qty[separate-uncertainty]{-1+-19}{\km\per\s}$ $\left[\Delta v=\qty[separate-uncertainty]{3+-34}{\km\per\s}\right]$. Thus, the systematic offset in wavelength calibration between the observations from NIRCam and NIRSpec is negligible.

In contrast, the best-fit centroids for Ly$\alpha$ and \heii\ $\lambda1640$ are redshifted relative to the systemic frame by roughly $250$ and $150$\,km\,s$^{-1}$, respectively. This behavior, well documented for Ly$\alpha$ at lower redshifts \citep[e.g.,][]{Trainor2015THEGALAXIES}, is commonly interpreted as a radiative-transfer effect in the presence of galactic outflows: gas at zero velocity or blueshifted relative to the systemic frame absorbs resonant photons, whereas emission from receding gas is redshifted out of resonance and preferentially escapes to the observer.

Recombination emission of \heii\ $\lambda1640$ is not subject to resonant foreground absorption to the same degree as Ly$\alpha$ and is an important diagnostic of the ionizing spectrum hardness. A redshift of \heii\ relative to systemic is marginally detected at $2.6\sigma$ significance in the {\tt MSAExp} reduction and $3.2\sigma$ in {\tt PypeIt}, likely driven by slight line asymmetry with enhanced wings to the red of the centroid. While the nearby \oiiisemi\ lines are narrow and unresolved in our G140M spectra, the \heii\ line is mildly broadened and slightly resolved with $\sigma_v\approx250$--$300$\,km\,s$^{-1}$.

Narrow \heii\ emission can arise in nebulae ionized by hot stars that produce the required 54 eV photons, whereas pronounced asymmetry or broadening is often interpreted as evidence for AGN or strong stellar winds \citep[e.g., Wolf--Rayet stars;][]{Leitherer2019HeReddening}. At the SNR of our stack, there is no unambiguous \heii\ P Cygni velocity signature indicating strong stellar winds, although a modest absorption component could produce the slight asymmetry seen here and would be clarified by data with higher-SNR or higher spectral resolution. At $\sigma_v\sim250$\,km\,s$^{-1}$ the \heii\ width is far narrower than typical broad-line AGN \citep[$\gtrsim2000$\,km\,s$^{-1}$;][]{Saxena2020TheSurvey} or Wolf--Rayet winds \citep[$\sim1000$--$3000$\,km\,s$^{-1}$;][]{Brinchmann2008GalaxiesUniverse}.  It could potentially reflect a contribution from low-luminosity narrow-line AGN, though we will argue below that this is slightly disfavored based on emission line ratios.

Our archival rest-frame optical EIGER spectra provide additional information via coverage of the \oiiiforbid $\lambda\lambda 4960,5008$\AA\ doublet, the associated $4364$\AA\ temperature-sensitive auroral line, and H$\beta$ as well as H$\gamma$.  Figure \ref{fig:NIRCam_masquerade_stack} shows a stack of these transitions as observed by NIRCam/WFSS for the MASQUERADE sample galaxies. The stack was constructed using the same numerical values as the rest-UV spectra in Figure \ref{fig:stairsreductions} to normalize the raw data and weight before averaging.  This ensures that each object contributes to the UV and optical stacks in the same proportion. Then, we fit the transitions following a comparable procedure to that for the absorption lines (Table \ref{tab:optical}).

\input{optical}

We considered the effect of systematic redshift uncertainties in the NIRCam data when constructing this stack, as differences have been reported between values for the same object measured from Module A vs Module B \citep{higginson}. 
These are at the level of $\Delta (z_A-z_B)=+0.0028$ or $+112$ \kms\ (systematic), $\pm0.0073$ or 292 \kms\ (random), likely caused by uncertainty in the source position centroid on the sky. In our stack, 40 galaxies are observed in both modules, so we average these as a reference zero-velocity rest-frame. Additionally, 24 galaxies are observed exclusively in Module A, and 40 exclusively in Module B; we can estimate the magnitude of any bias in the stacked velocity centroids from this imbalance by noting that each individual module is offset by $\sim 56$ \kms\ (half the systematic difference), and that the stack's total offset is therefore of order $56\times (24-40)/104=-8.6$ \kms\ toward Module B's zero point with an uncertainty of $3.6$ \kms. We do not attempt to correct for this offset, as it is smaller than the stack's individual pixels, and will be even smaller when considering the inverse variance weighting scheme used, which will (on average) favor the galaxies observed in both modules.  

In the NIRCam stack, we observe tentative evidence in the \oiiiforbid profile of velocity wings that are distinct from the central core of the emission line. A two-component Gaussian model converges on a line core with $\sigma=\qty[separate-uncertainty]{128+-7}{\km\per\s}$ and a weaker component capturing broad wings with $\sigma={351}_{-22}^{+24}\;\mathrm{km\,s^{-1}}$. Broad wings in the \oiiiforbid profile have been previously interpreted as a signature of galactic outflows \citep{oiii_emission_outflows,emission_outflows_xu}. Again, these velocities could be characteristic of star-formation driven outflows or weak narrow-line AGN. However, we are mindful that the data processing in EIGER includes a step of high-pass filtering that can affect the detailed profiles of emission lines in NIRCam spectra \citep{EIGER_I}. Furthermore, any spatial extent along the dispersion's direction will also extend the lines. We examined a separate set of NIRCam reductions designed to mitigate filtering artifacts \citep{2026A&A...706A.165K} without finding any qualitative difference or strengthened evidence. We therefore state conservatively that the emission profile is {\em consistent} with emission from winds, but the evidence on this alone is not as strong as what is described below based on absorption signatures.

\subsection{Emission Line Ionization Diagnostics}

The combined UV and optical spectral coverage in these stacks yields a number of line ratios that can be used to study ionization conditions.

First, a clear detection of strong \oiiiforbid $\lambda\lambda 4960,5008$\AA\ and H$\gamma$ without associated \oiiisemi $\lambda 4364$\AA\ provides evidence that the stack is dominated by an ISM with low to moderate electron temperature $T_e$. While \citet{Matthee2023EIGER.JWST} report a $5.3\sigma$ detection of \oiiiforbid $\lambda 4364$\AA\ in a stack of all 117 galaxies in the J0100 field, the full MASQUERADE sample---which has 104 objects; spans a second field; normalizes by UV $f_\nu$ rather than $f_\lambda$ around [\ion{O}{3}] 4960, 5008; and is selected for higher continuum luminosity---shows no evidence of auroral line emission.  This is reflected in the ratio of H$\gamma$/\oiiiforbid$_{4364}$, which Matthee et al. report as 2.125, but we measure at $\gtrsim 3$ ($3\sigma$ lower limit). More directly, using a measured upper limit on $I_{4364}$, we find a $3\sigma$ lower limit on the ratio:
\begin{equation}
R_{\rm [OIII]}\equiv\frac{I(4960+5008{\rm \text{\AA}})}{I({4364}{\rm \text{\AA}})}\gtrsim 89.
\end{equation}
Together with a low dust reddening and attenuation inferred from the ratio $I({\rm H}\beta)/I({\rm H}\gamma)=2.27\pm0.32$ (consistent with the Case B expectation of 2.14), $R_\mathrm{[O\;\textsc{iii}]}\gtrsim89$ implies an excitation temperature $T_e\lesssim 13,600$K, which disfavors a dominant ionization contribution from AGN.

On the other hand, the same ionization conditions can be produced by very young stellar populations with hard ionizing continua (e.g., Wolf--Rayet stars, stripped-envelope binaries, X-ray binaries, or hot, massive low-metallicity stars). These scenarios are all plausible in the early universe and are not easily distinguished from narrow-line AGN at our spectral resolution. The H$\alpha$ region of the classic optical BPT diagram lies outside of our NIRSpec/NIRCam spectral coverage, but we measure a stacked value of
\begin{equation}
{\rm [O\ III]_{5008}}/{\rm H}\beta=9.9\pm1.2,
\end{equation}
\noindent which is near the maximum value expected of either star forming galaxies or AGN \citep{kewley_ionization}, and $\approx 3\times$ higher than typical low-redshift star forming galaxies. However at $z\sim 6$ our spectra do not cover [N II] or H$\alpha$, and the locally-derived BPT diagram becomes a less discriminating diagnostic as lower-metallicity star forming galaxies approach the AGN locus, and metal-poor narrow-line regions also move into the star-forming locus \citep{2023A&A...677A.145U,2023ApJ...959...39H,2025A&A...697A.175S}.

For these reasons, several alternative rest-UV ratios have been proposed as AGN/star-formation discriminants for high redshift \citep{feltre_2016}. We derive the following values from the {\tt MSAExp} measurements in Table \ref{tab:linemeasurements}:
\begin{eqnarray}
    \mathrm{O\;\textsc{iii}]}_{1666}/\mathrm{He\;\textsc{ii}}_{1640}=0.58\pm0.18\\
    \mathrm{C\;\textsc{iii}]}_{1909}/\mathrm{He\;\textsc{ii}}_{1640}=1.31\pm0.33\\
    \mathrm{C\;\textsc{iv}}_{1548}/\mathrm{He\;\textsc{ii}}_{1640}=0.67\pm0.22
\end{eqnarray}

Comparing these ratios to star-forming and AGN photoionization models places the stack in a region compatible with highly ionized outliers among young starbursts, while still lying near the overlap with low-ionization AGN. The observed \ciiisemi/\heii\ ratio in particular is difficult to reproduce with AGN models \citep{feltre_2016}, but this diagnostic should be treated cautiously given the potential second-order contamination in the \ciiisemi\ spectral region.  

\subsection{Emission Lines' Summary}
Summarizing our results on the velocity profiles and ratios of stacked emission-lines, we find:
\begin{itemize}
    \item The semiforbidden lines are narrow and agree well with the emission-line redshifts obtained by EIGER using NIRCam slitless spectroscopy, validating cross-instrument wavelength and redshift calibration.
    \item The Ly$\alpha$ profile has similar strength to $z\sim3$ stacks and is redshifted by $\sim250$\,km\,s$^{-1}$ relative to the systemic frame.  Likewise there is tentative evidence of broadening in the \oiiiforbid profile. We interpret these as velocity signatures of outflows.
    \item The presence of \heii, \oiiisemi, and \ciiisemi, together with high [O\,\textsc{iii}]/H$\beta$ and non-detection of [O\,\textsc{iii}] $\lambda4364$, indicates an enhanced ionization condition but a low electron temperature in the composite spectrum of the ISM. The high-ionization lines appear stronger than in $z\sim3$ stacks.
    \item The narrow emission line widths favor ionization driven by young, hot stars. The linewidths rule out a dominant contribution from broad-line AGN, but cannot strictly exclude narrow-line nuclear emission.
\end{itemize}

Together, these diagnostics paint a picture of an ISM that is highly ionized by sources that have significant flux at the \heii\ edge but are not energetic enough to drive heating that would produce auroral emission or generate large velocity widths. The required stars have $T_{\rm eff}\approx 60,000-100,000$K (hotter than the O sequence, which is opaque at the He+ edge) and could include binary stripped Helium stars \citep{gotberg}, or low-metallicity massive stars in binary population models \citep{stanway_eldridge, schaerer2002}.

The evidence therefore favors an interpretation where ionization in the composite is driven by similar hot stars rather than AGN, to the extent that these can be disambiguated given our SNR and spectral resolution plus the reliability of model diagnostics.

\subsection{Absorption Line Properties\label{subsec:absorptionproperties}}

With the rest frame cross-validated between NIRSpec and NIRCam, we now turn to absorption lines, many of which show clear evidence for a blueshift relative to the systemic rest frame.

Figure \ref{fig:emline_offsets} provides a visual summary of the centroid measurements in Table \ref{tab:linemeasurements}, showing broad agreement around a $\Delta v\sim-200$\,km\,s$^{-1}$ velocity offset for \siii, \cii, \oi, \siiv, and \civ. The \nv\ doublet exhibits an even larger blueshift, but we discard it from subsequent analysis because the continuum level in that region is compromised by proximity to Ly$\alpha$ in emission and absorption. \feii is the weakest line detection and is slightly offset from the mean value for the {\tt MSAExp} reduction, but not at a statistically significant level. Including all lines except \nv, we find a weighted-average velocity offset of \qty[separate-uncertainty]{-178+-36}{\km\per\s} at $4.9\sigma$ significance [\qty[separate-uncertainty]{-204+-35}{\km\per\s} at $5.8\sigma$ significance].  

Although the \civ and \siiv lines are among our cleanest absorption fits, they are also more likely to reflect stellar wind signatures distinct from any outflows from the bulk ISM \citep[e.g.][]{2018ApJ...853...87R}.  The detailed \civ profile in Figure \ref{fig:plottransitions} in particular shows evidence of residual absorption at larger $\Delta v$ not captured by our simple Gaussian fit, at velocities up to $\sim\qty{1000}{\km\per\s}$ that are characteristic of stellar winds.  If one excludes these transitions from the average to focus only on low-ionization ISM lines, the resulting outflow velocities are reduced to $-166\pm 49$ \kms [$-178\pm 46$ \kms].  

\section{Discussion}

\subsection{Outflow Trajectories}
\label{sec:trajectories}

Our blueshifted absorption-line measurements provide statistically significant evidence for outflows that transport enriched gas out of star-forming regions, qualitatively similar to winds observed in starburst galaxies at lower redshift. Given that heavy-element absorption is routinely seen in quasar spectra even at these redshifts \citep{davies_xqr30}, and that galaxies in the first Gyr are typically less massive than their later counterparts, it is tempting to link these outflows to intergalactic and circumgalactic chemical enrichment. However, this connection must be balanced against the short Hubble time and the still-shorter star-formation ages of these systems, which may limit the distances winds can traverse.

To assess wind propagation we integrate a toy model in which a wind particle moves under the exclusive influence of the galaxy's gravity (i.e., neglecting hydrodynamical interaction with the IGM or nearby companions). Expressed in comoving distance $\chi$, the equation of motion is:
\begin{equation}
    \ddot{\chi} = -2H\dot{\chi} - \frac{GM}{a^3\chi^2}\label{eq:chronology}
\end{equation}
where $H=H(z)$ is the Hubble parameter, $a$ is the scale factor, and $M$ is the enclosed mass (assumed to be dark-matter dominated). The second term on the right-hand side is the galaxy's gravitational acceleration, while the first term captures Hubble drag on peculiar velocities; this drag is larger at high redshift [e.g., $H(z=6.5)\approx778$\,km\,s$^{-1}$\,Mpc$^{-1}$], but still a smaller effect than gravity because winds start deep within the halo potential.

We integrate Equation \ref{eq:chronology} for every galaxy in the sample over the age of its stellar population, assuming a fixed initial outflow velocity $v_{\rm outflow}=200$\,km\,s$^{-1}$ that we found in Subsection \ref{subsec:absorptionproperties} plus initial positions and masses derived from NIRCam photometry (described below). This represents the maximum distance a cosmological and free-falling wind could travel if launched at the onset of star formation.

We set the initial position of each wind particle at the galaxy's Kron radius converted to physical units, measured from F356W photometry using {\tt SExtractor} \citep{Matthee2023EIGER.JWST} and the angular diameter distance appropriate to that object's $z$. This choice implicitly assumes order-unity covering factors for the absorbing material. Galaxies at these redshifts are compact; for our sample, the median Kron radius is $0.\!^{\prime\prime}376$, corresponding to 2.18 proper kpc. Depending on morphology, the Kron radius is typically $2$--$3\times$ larger than the half-light radius $R_e\sim1$\,kpc. Launching winds from the Kron radius rather than from $R_e$, and also using an initial velocity of $-200$ \kms\ for all ions rather than $-170$ \kms\ for the low-ions alone, yields larger final wind bubbles and therefore provides a conservative upper bound on enrichment radii.

\begin{figure}
    \plotone{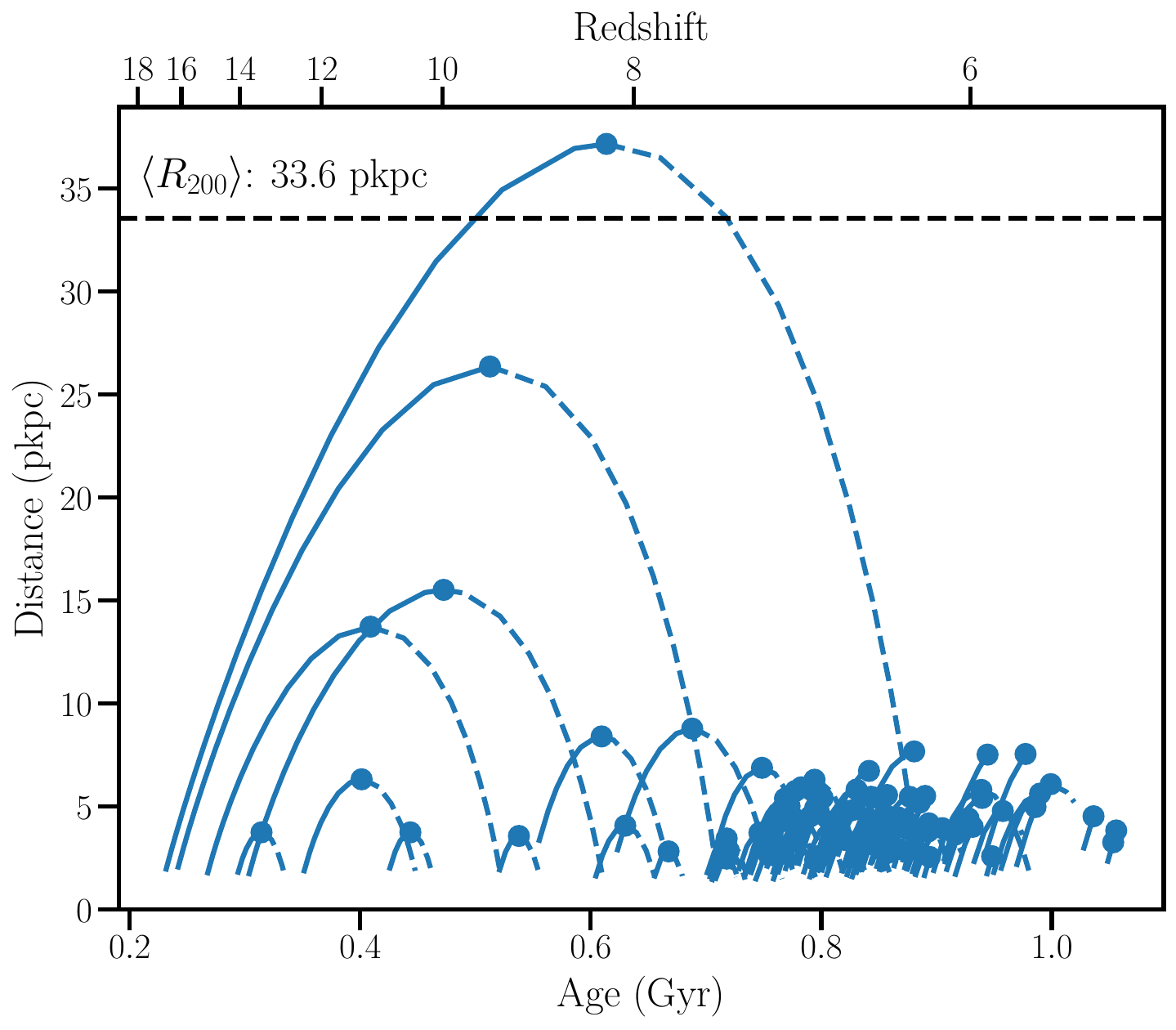}
    \caption{Summary of wind trajectories for the sample galaxies (see text). Solid blue lines indicate outflowing portions of the trajectories, while dashed lines depict turnaround and infall where applicable. The solid blue dot on each trajectory marks its maximum achieved wind distance.  For young winds this is bounded by the integrated distance over travel time, but for old gravitationally bound winds the maximum is the turnaround radius. With a dashed black line, we demarcate the average $\langle R_{200}\rangle$ for our halos according to \texttt{MSAExp}'s weights. $\langle R_{200}\rangle$ for \texttt{PypeIt} is similar: \qty{33.7}{pkpc}.}
    \label{fig:wind_trajectories}
\end{figure}

{\tt Prospector} fits to the NIRCam data (which incorporate photometry in F115W, F200W, and F356W as well as line fluxes for \oiiiforbid $\lambda\lambda 4960, 5008$ and H$\beta$) also provide a posterior on each galaxy's age $t_{20}$ (the time at which 20\% of each galaxy's total stellar mass was in place). We set the wind launch time for each galaxy to this epoch, effectively integrating for 80\% of its inferred star-formation history.

The {\tt Prospector} fits consist of a composite SED with a stellar population plus nebular emission lines calculated using {\tt CLOUDY} model grids. Explicit inclusion of \oiiiforbid and H$\beta$\ line fluxes in the fits was essential to obtain meaningful results, as these lines sometimes dominate the total flux in broadband photometric filters. We experimented with delayed-$\tau$ and non-parametric models of the star formation history (SFH). Because most galaxies are young by construction at $z\sim 6$, the delayed-$\tau$ model was a less useful representation of its SFH: essentially every galaxy is in a quasi-linear rising phase of the SFH.  We therefore quote model fits based on {\tt Prospector}'s non-parametric SFH fits. Given our small compliment of filters, the derived stellar mass is the most robust output parameter, with SFR and galaxy age carrying larger uncertainty (these parameters are coupled for a given stellar mass). We do not use the derived SFR for analysis, but $t_{20}$ is an important input to wind models. The posterior distribution of {\tt Prospector}'s $t_{20}$ quantitatively encodes this age uncertainty, which we propagate into the wind model and consider carefully when interpreting model results.

For the halo mass in Equation \ref{eq:chronology}, we apply a time-evolving NFW profile. From \texttt{Prospector}'s inferred stellar mass, we derive the halo mass via abundance-matching according to \texttt{UniverseMachine} \citep{2019MNRAS.488.3143B}. Setting this as the halo mass at the epoch of observation, we backtrack the prior halo mass at earlier times using fitting formulae for the median halo growth in the Millennium simulation \citep{halo_growth}.

We then integrate Equation \ref{eq:chronology}, evaluating the scale factor, Hubble parameter, and the evolving halo mass at each time/redshift step. Given the redshift and halo mass, we query {\tt COLOSSUS} \citep{Diemer2018COLOSSUS:Halos} for the NFW profile's concentration at each step. Only the mass interior to the instantaneous wind-particle radius contributes to the gravitational term.

To gauge how uncertainties in $t_{20}$ and the halo mass affect our radii of enrichment, we run three scenarios. In the first ``median" scenario, we set $t_{20}$ and the halo mass at their posterior distributions' medians. We then run a ``maximal" or conservative wind scenario that maximizes the radius of enrichment by employing the eighty-fourth percentile of $t_{20}$ to extend the outflow's time and the halo mass matching the inferred stellar mass's sixteenth percentile to minimize gravitational deceleration by the wind. Our third run is a Monte Carlo simulation. For every galaxy, we draw and integrate 100 random samples from posterior distributions of $t_{20}$ and the halo mass. 

\begin{figure}
    \epsscale{1.15}\plotone{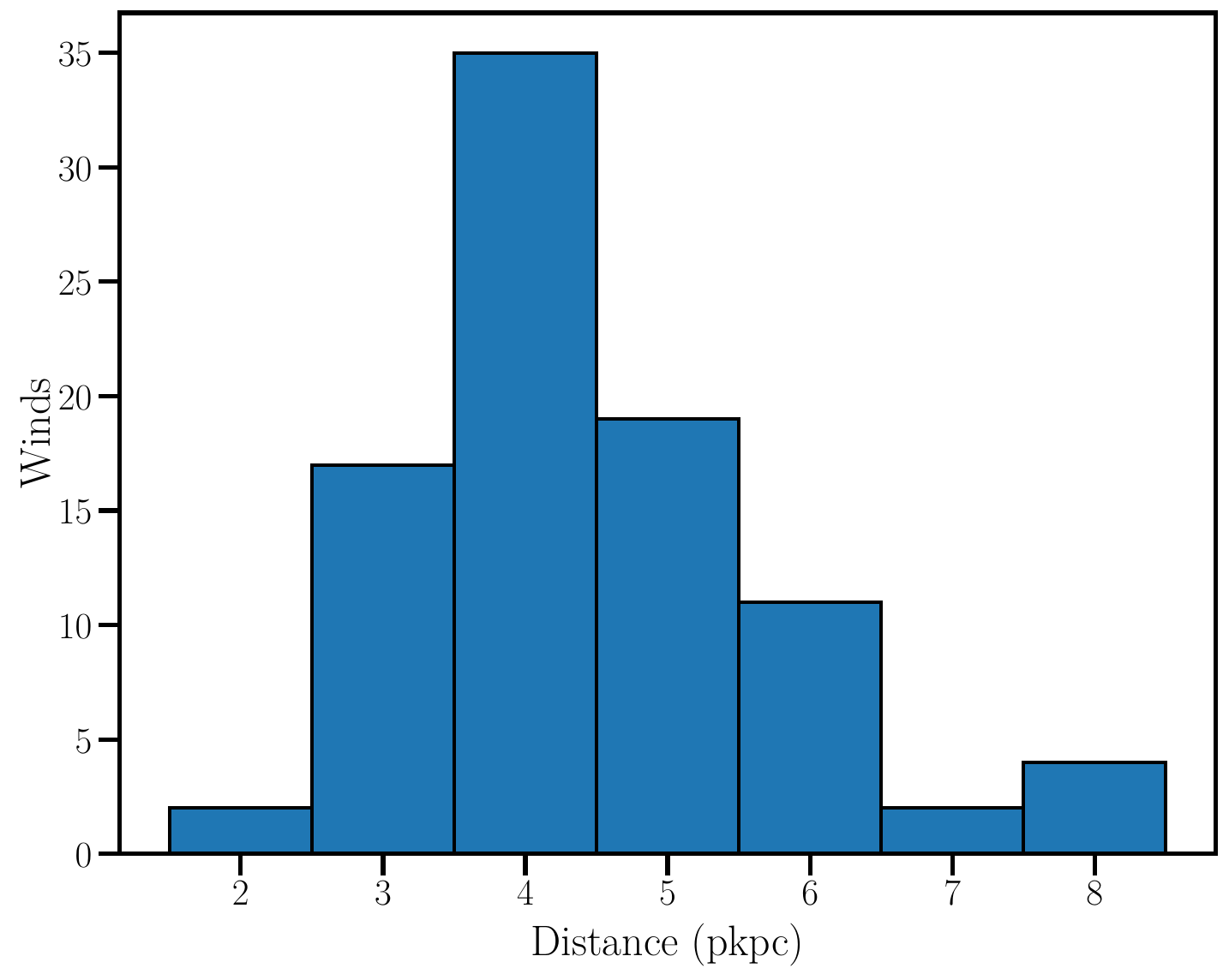}
    \caption{Histogram of maximum wind distances $D_{\rm max}$ for our median scenario. $D_{\rm max}$ is either the turnaround radius for stalled winds or the propagation-limited radius at the observation epoch. For compactness, this histogram excludes the four longest-distance winds: \qty{13.7}{kpc}, \qty{15.2}{kpc}, \qty{26.4}{kpc}, and \qty{37.2}{kpc}.}
    \label{fig:wind_histogram}
\end{figure}

Figure \ref{fig:wind_trajectories} shows individual wind trajectories in our median scenario. The majority of the sample is very young ($t_{20,\mathrm{med}}\approx\qty{38}{Myr}$), so a 200\,km\,s$^{-1}$ wind launched at the onset of star formation will have traveled only a few proper kpc and not yet reached any terminal radius. For some older systems, the wind stalls at $<10$\,kpc and material falls back; this simple model does not include later episodes of star formation and recurrent outflows that can cycle material between the ISM and the emerging CGM. For the 12 galaxies with implied formation redshifts $z>8$, a 200\,km\,s$^{-1}$ wind launched from $\lesssim2$\,kpc in a $10^{10}\,\mathrm{M_\sun}$ halo remains bound, reaching a maximum distance $D_{\rm max}\sim5$--$15$\,kpc before stalling, though two galaxies do approach the virial radius. The large majority begin forming stars at $z\approx6.5$--$7.0$ and have ages $<100$\,Myr; for all but the most massive halos, their winds have not yet stalled but also have not had time to propagate more than $D\sim5$--$10$ proper kpc.

These results are summarized in the histogram of $D_{\rm max}$ in Figure \ref{fig:wind_histogram}. The largest wind bubble, not shown in Figure \ref{fig:wind_histogram} for compactness, implied by this exercise is $\sim37$\,kpc, comparable to the virial radius $R_{200}$ for our average sample galaxy. Most objects yield $D_{\rm max}\approx3$--$9$\,kpc, meaning they can lie well outside the stellar component but may not yet have escaped their dark-matter halos into the IGM.

\begin{figure}
    \epsscale{1.15}
    \plotone{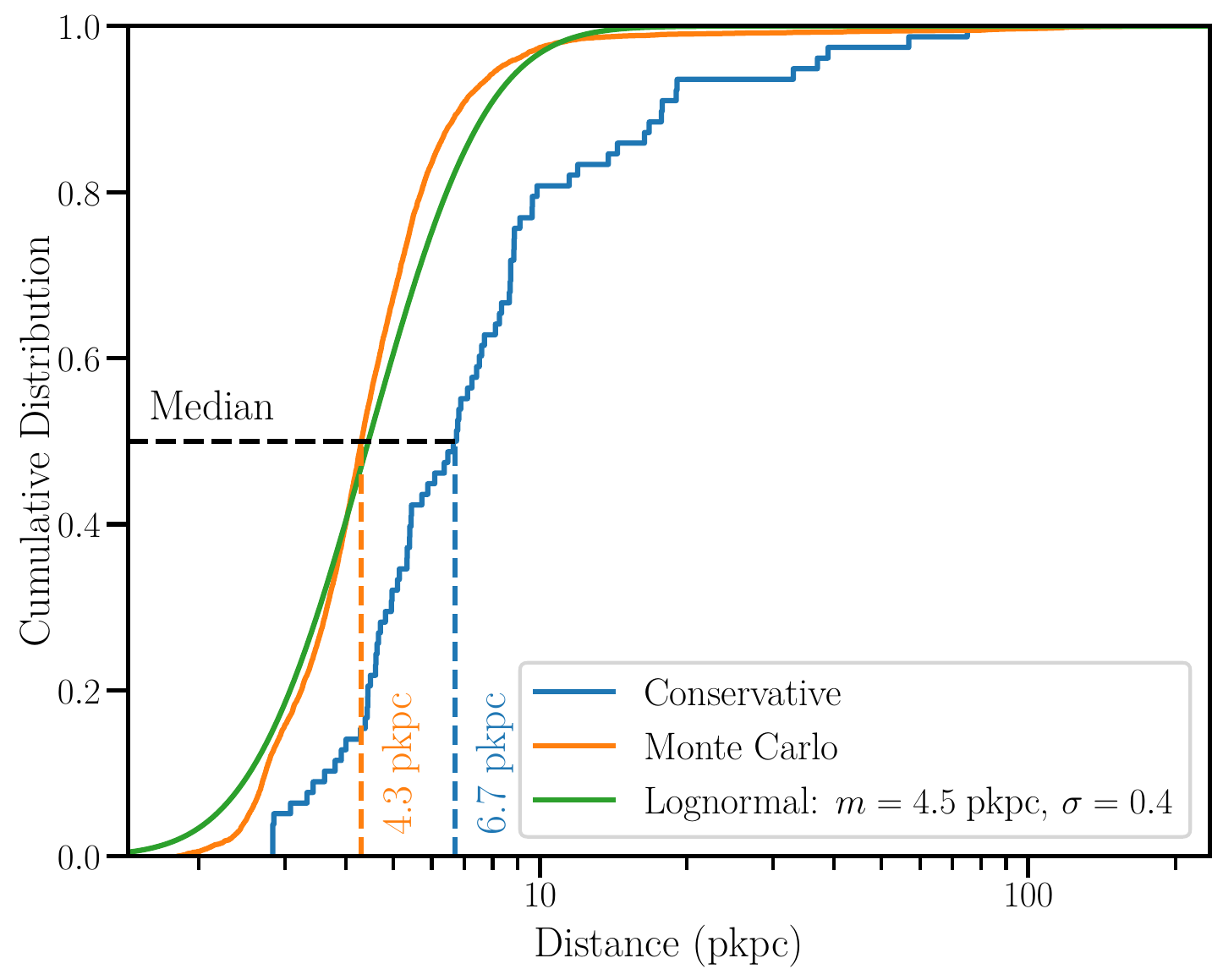}
    \caption{Empirical cumulative distribution functions for our conservative and Monte Carlo scenarios. We overlaid a lognormal fit to the Monte Carlo distribution.\label{fig:conservativemontecarlo}}
\end{figure}

Figure \ref{fig:conservativemontecarlo} depicts cumulative distribution functions for our conservative and Monte Carlo scenarios. We also marked each distribution's medians and fit a lognormal distribution to the Monte Carlo scenario. The Monte Carlo scenario's median of \qty{4.3}{pkpc} approximately matches that of the median scenario. While the conservative scenario boosts the median distance by around 55\%, the distances remain small, as only about 20\% reach \qty{10}{pkpc}. Altogether, the scenarios convey that our finding of short-distance winds from the onset of star formation until the epoch of observation is robust to inferential uncertainties.

\subsection{Implications for Production of Metal Absorption in Quasar Sightlines}

Early JWST observations of galaxies in quasar fields have yielded several examples of galaxy/metal-absorber pairs at nearly coincident redshift \citep{higginson, bordoloi24, zou_aspire, aspire_galabs}. Here we examine whether the winds observed in MASQUERADE's [O\,\textsc{iii}] emitters could contribute meaningfully to the total cross section of randomly intervening heavy-element absorption systems.

In the EIGER quasar fields, \citet{higginson} report five groups of [O\,\textsc{iii}] emitters near neutral O\,\textsc{i} absorbers, with the nearest galaxy to the sightline having impact parameters $b=25,\,41,\,120,\,148,$ and $158$\,pkpc, respectively. Likewise, \citet{bordoloi24} examined Mg\,\textsc{ii} absorbers in one of the observed EIGER fields over a wide redshift range. If we consider galaxies at $z>5.7$ from that sample (i.e., matched in redshift to the MASQUERADE stack), the nearest object to the sightline is at $b=82$\,pkpc (excluding a galaxy--absorber pair at 17\,pkpc at the redshift of the background quasar), and the remaining objects lie at $120<b<250$\,pkpc.

The maximum wind radii derived from our toy model---which assumes an initial 200\,km\,s$^{-1}$ outflow velocity launched at the Kron radius and ages inferred from stellar population fits---are too small to reach these impact parameters. This suggests that the metals observed as absorption in the quasar spectra may not normally originate from the nearest observed galaxy and therefore cannot be immediately interpreted as its own circumgalactic matter. 

The principal caveat to this conclusion is that MASQUERADE survey galaxies are brighter (by design) than randomly selected counterparts from blind surveys like EIGER. The average object in \citet{higginson} or \citet{bordoloi24} has an order-of-magnitude lower stellar mass which corresponds to a $\sim 3\times$ reduction in halo mass decelerating the wind. Integrating Equation \ref{eq:chronology} with a $3\times$ lower halo mass reveals that the final radius in an NFW halo depends on $M_h$ to a greater degree than a point mass but still with a fairly shallow slope. The maximum radius increases by 50\%, which is significant but not sufficient to bridge the gap between inferred and observed impact parameters. While a reduction in mass alone is not sufficient to launch winds to large distance for any single halo, a combination of slightly larger wind bubbles and increased number density of fainter galaxies can increase the probability of coincident identifications.

We can interrogate this hypothesis statistically by estimating the number density of absorption systems that observed [O\,\textsc{iii}] emitters could plausibly produce in a survey like EIGER or MASQUERADE (as MASQUERADE is a subset of EIGER). Assuming winds from [O\,\textsc{iii}] emitters are metal enriched and contribute to the intervening metal lines seen in QSO spectra, the absorption frequency can be estimated via the cross section $\sigma=\pi R_{\rm max}^2$ (in physical units) and the comoving number density of host galaxies $n_c$ derived by integrating the luminosity function:
\begin{equation}
    \frac{dN}{dX} = \frac{c}{H_0}\,n_c\sigma\label{eq:absorberdistributionfunction}.
\end{equation}
Integrating the luminosity function from \citet{Matthee2023EIGER.JWST} --- the parent sample for MASQUERADE ---
over $\log{\left[L_{[\mathrm{O\;\textsc{iii}}]}/\left(\mathrm{erg\,s^{-1}}\right)\right]}=[42.0,43.4]$ with $\alpha=-2.0$ yields $n_c=0.00152$\,cMpc$^{-3}$. We can then simplify Equation \ref{eq:absorberdistributionfunction}:

\begin{equation}
    \left.\frac{dN}{dX}\right|_\mathrm{[O\;\textsc{iii}]\ emitters}= 0.00095 \left(\frac{R_{\rm max}}{6.7~ {\rm pkpc}}\right)^2.
\label{eq:o3emitters}
\end{equation}

\noindent where the reference value of 6.7 pkpc reflects the median of the conservative (in the sense of large wind distances) distribution in Figure \ref{fig:conservativemontecarlo}. If we instead use the Monte Carlo distributions, we find $dN/dX={6.2}_{-4.4}^{+3.9}\times{10}^{-4}$.

There are numerous surveys of heavy-element absorbers in high-redshift quasar spectra that report $dN/dX$ values much larger than these values. For example, \citet{davies_civ} find $dN/dX_{\rm CIV}=0.16\ (0.79)$ for absorbers with $13.8<\log(N_{\rm CIV})<15.0$ ($13.2<\log(N_{\rm CIV})<15.0$). \citet{becker_oi} studied O\,\textsc{i} absorbers with sensitive equivalent-width limits ($W>0.05$\,\AA) and reported $dN/dX_{\rm OI}=0.421^{+0.098}_{-0.101}$. \citet{chen_z6_mgii} surveyed 100 sightlines for Mg\,\textsc{ii} at similar redshifts and $W>0.3$\,\AA, finding $dN/dX_{\rm MgII}=0.324\pm0.042$, and \citet{christiansen_metals} report comparable values for Mg\,\textsc{ii}, C\,\textsc{ii}, and Si\,\textsc{ii} (with differing equivalent-width thresholds across surveys).

It is neither straightforward nor particularly informative to attempt a detailed reconciliation of differing ion-specific $dN/dX$ measurements and equivalent-width cutoffs here. Instead, we emphasize that our toy-model wind bubbles under-produce the observed heavy-element QSO absorption frequencies by a factor of $\sim 160$--$4500$ for plausible choices of ion and sensitivity threshold.

The key implication is that the galaxies represented in MASQUERADE's stack are generally too young, and their winds are too slow, to pollute enough volume to produce most of the metal lines seen in the IGM. Reconciling this discrepancy requires either (1) substantially larger wind bubbles, or (2) enrichment driven by fainter galaxies clustered near the observed [O\,\textsc{iii}] emitters, or both. The fainter EIGER sample, however, has uncovered very few \oiiiforbid emitters at impact parameters approaching our bubble radii, nor even fainter photometric candidates.

Matching intervening absorber observations via bubble cross section alone would require winds to travel a larger distance, by a factor equal to the square root of the observed-to-inferred $dN/dX$ ratio. If this ratio is $\sim160$--$4500$, then bubble radii would need to be $\sim13$--$67\times$ larger. Factors that influence the propagation distance in our toy model are galaxy age and mass (derived from stellar-population synthesis and halo abundance matching), as well as initial wind velocity; we consider each in turn below.

{\it Age:} From Figure \ref{fig:wind_trajectories}, it is clear that most wind bubbles are limited by finite propagation time more than by gravitational deceleration, indicating that galaxy age is a key determinant of bubble size. In our {\tt Prospector} fits based on only three photometric bands and two independent line fluxes (see Section \ref{sec:trajectories}), stellar mass is the most robust parameter, while age and SFR are somewhat more uncertain. The median {\tt Prospector} $t_{20}$ for our sample and integration time for the wind model is \qty{38}{Myr}. For our median galaxy with $M_{\star}=10^{8.35}\;\mathrm{M_\sun}$, the expected SFR on the main sequence or starburst track is $\sim$1--10\;$\mathrm{M_\sun}$\,yr$^{-1}$; we find a median SFR of $3.8\;\mathrm{M_\sun}$\,yr$^{-1}$ averaged over \qty{10}{Myr} from {\tt Prospector}, and a consistent $3.6\;\mathrm{M_\sun}$\,yr$^{-1}$ when scaled from the H$\beta$ flux. Dividing stellar mass by SFR yields an order-of-magnitude age consistent with the {\tt Prospector} outputs. Thus, while older galaxies could produce larger bubbles, doing so would typically require SFRs below expectations for the main sequence and weaken the winds. Errors in ages or bubble cross sections are unlikely to resolve the orders-of-magnitude discrepancy.

Further support for young star formation ages comes from independent {\tt beagle} modeling of UV-bright galaxies like those in MASQUERADE but with many more photometric bands \citep{2025ApJ...987..189E}. For objects at $M_{\rm UV}\sim -20$, the majority of their inferred stellar mass is generated in recent bursts with ages of tens of Myr and rising SFR. 

{\em Initial Velocity:} Galactic wind turnaround radii are sensitive to the initial wind speed in our toy model, which we set to $-200$ \kms\ to match our fits to absorption velocity centroids ($\sim 35$ \kms\ uncertainty). The absorption fits are well converged, and it is unlikely that these Gaussian profiles (essentially an optical-depth weighted velocity) are incorrectly fitting material that is moving faster by the factor of $13-67$ necessary to explain the metal-line $dN/dX$.  

However, in many lower redshift outflows, one observes an asymmetric absorption profile characterized by a higher velocity tail extending to the blue of the line center, particularly in high-ionization species. We see tentative evidence of this in a broadening of the \civ profile, but, as noted elsewhere \citep[e.g.,][]{2018ApJ...853...87R, 2023MNRAS.526.6297S, 2016MNRAS.458.1891B}, \civ absorption can arise from fast stellar winds that can be difficult to disentangle from the outflowing ISM. We do not observe high-velocity tails in other ions. Although we acknowledge the possibility of fast winds not visible at the SNR of our stack, the present data do not show compelling evidence of asymmetry in the low-ionization lines that are most clearly associated with galactic outflows. 

{\em Galaxy Mass:} We explained above that uncertainties in stellar mass estimates have only a weak effect on outflow distance for any individual galaxy and cannot explain the discrepancy in $dN/dX$. However, when considering the collective galaxy population, an alternative route is enrichment from fainter, undetected galaxies that are more numerous by 2--3 orders of magnitude. For a faint-end slope $\alpha=-2$ of the UV luminosity function \citep{Bouwens2021NewEfficiency, Atek2026AModels}, the number density increases by one dex for each dex fainter in luminosity. Hence, it would be feasible to reproduce the observed $dN/dX$ if winds of similar radius originate from objects $2.2$--$3.7$ dex fainter than those in our MASQUERADE stack (i.e., $M_{\star}\sim10^{5}$--$10^{6}\;\mathrm{M_\sun}$). 
\citet{2025ApJ...987..189E} found that these lower mass galaxies often formed more of their stellar mass at earlier times, thereby providing more time for winds to propagate. On the other hand, \citet{2025ApJ...992..191G} measure weaker outflow velocities in their stack of lower-mass galaxies than we do for MASQUERADE.

A detailed quantitative analysis of these interacting effects is beyond the scope of this paper; most likely a combination of all the above is at play. While bright galaxies like those in our MASQUERADE stack almost certainly do not produce the majority of QSO metal absorbers, they may account for a few percent of the total and are likely to cluster with satellites that have $10$--$100\times$ lower stellar mass. These smaller objects are $10$--$100\times$ more numerous and can provide a larger aggregate cross section. The absorber--[O\,\textsc{iii}] emitter pairings seen at $b\gg R_{200}$ in \citet{bordoloi24} and \citet{higginson} may reflect clustering on group or larger scales (Simcoe, in prep.), with dynamical interactions helping to distribute mass and metals in ways not captured by the simple models considered here.

\section{Conclusions}

We present a stack of 104 galaxies with $5.7 < z < 6.9$ and F115W $<27$, observed with JWST/NIRSpec's G140M grating and covering rest-frame UV wavelengths from Ly$\alpha$ through \ciiisemi\ $\lambda1909$\,\AA\ at $R\sim1000$. The sample has a weighted mean redshift of $z=6.50$ and mean stellar mass $M_\star=10^{8.93}\,M_\sun$. The stack has continuum SNR of 10.25 (7.83), which enables detailed comparison of emission and absorption line properties with other composite spectra in the literature at varying redshifts. Our main findings are:
\begin{enumerate}
    \item Qualitatively, the normalized stack is extremely similar to the $z\sim3$ composite of Lyman-break galaxies by \citet{shapley} in continuum shape, emission, and absorption line properties, despite differences in selection (blind [O\,\textsc{iii}] grism survey versus Lyman break selection).
    \item We observe strong, unresolved emission from \oiiisemi\ $\lambda\lambda1660,1666$, \ciiisemi\ $\lambda1909$, and marginally resolved \heii\ $\lambda1640$; all but \heii\ are at systemic velocities consistent with the [O\,\textsc{iii}] rest frame. These three lines are stronger in our stack than the stack by \citet{shapley} at $z\sim3$, implying higher ionization at earlier times. At NIRSpec G140M resolution, we can exclude a dominant contribution from broad-line AGN; narrow-line AGN remain challenging to rule out for the narrowest features, and higher-resolution data would help distinguish between an ISM dominated by ionizing radiation from starbursts, and one influenced by a low-luminosity, narrow line AGN.
    \item If the UV emission lines are excited by stellar populations, those populations must include young, hot stars capable of producing photons that ionize He$^{+}$. The observed equivalent-width ratios in this interpretation are challenging to reproduce using standard stellar populations.
    \item We detect clear blueshifted absorption lines from low- and high-ionization tracers of stellar and/or galactic winds. MCMC fits yield mean velocity offsets of $\sim170$--$200$\,km\,s$^{-1}$ relative to systemic.
    \item A simple kinematic model indicates that a 200\,km\,s$^{-1}$ wind propagating for a typical integration time $t_{20}\sim\qty{38}{Myr}$ will reach $R_{\rm max}\sim3$--$9$\,pkpc. Most galaxies have $R_{\rm max}$ limited by travel time; a few reach a terminal radius and remain bound, falling back toward the center.
    \item We estimate the contribution of [O\,\textsc{iii}] emitters to the absorber incidence $dN/dX$ and find that $R_{\rm max}\sim5$\,kpc wind bubbles from these galaxies can account for less than a percent of observed absorption systems in quasar spectra.
    \item While uncertainties in ages and propagation times could increase some galaxies' cross sections, the combination of large observed absorber--galaxy impact parameters and our modeled underproduction of absorbers suggests that most metal enrichment arises from undetected objects with $M_{\star}\sim10^{5}$--$10^{6}\,\mathrm{M_\sun}$ clustered around observed galaxies.
\end{enumerate}

Overall, the composite spectrum shows remarkably little evolution between $z\sim3$ and $z\sim6.5$, aside from stronger high-ionization emission lines in the higher-redshift stack. Higher spectral-resolution observations of these lines could more decisively rule out narrow-line AGN if the lines remain unresolved. Our results add to mounting evidence that, while emission-line galaxies observed with JWST are correlated with the sites of early enrichment, the bulk of intergalactic metals at these redshifts likely originate in fainter, more numerous systems clustered around the objects we detect.

\begin{acknowledgments}
This work is based on observations made with the NASA/ESA/CSA James Webb Space Telescope. The data were obtained from the Mikulski Archive for Space Telescopes at the Space Telescope Science Institute, which is operated by the Association of Universities fro Research In Astronomy, Inc. under NASA contract NAS 5-03127 for JWST.  These observations are associated with programs \#1243 and \#4713.  The specific observations can be accessed via doi:10.17909/w7hm-qb39.

We acknowledge the Massachusetts Institute of Technology's Office of Research Computing and Data for managing the Engaging cluster computer on which this article's research was performed. JM and GK acknowledge funding by the European Union (ERC, AGENTS, 101076224).
\end{acknowledgments}
\facility{JWST (NIRCam and NIRSpec)}
\software{Astropy \citep{TheAstropyCollaboration2022ThePackage}, COLOSSUS \citep{Diemer2018COLOSSUS:Halos}, emcee \citep{Foreman-Mackey2013EmceeHammer}, Matplotlib \citep{Hunter2007Matplotlib:Environment}, MSAExp \citep{msaexp, msaexp_science, msaexp_heintz}, NumPy \citep{Harris2020ArrayNumPy}, PypeIt \citep{Prochaska2020PypeIt:Pipeline}, schwimmbad \citep{Price-Whelan2017Schwimmbad:Python}, SciPy \citep{Virtanen2020SciPyPython}, and Specutils \citep{Earl2025Astropy/specutils:V2.2.0}}
\bibliographystyle{aasjournalv7}
\bibliography{stack}

\appendix
\section{Figures According to \texttt{PypeIt}}
\begin{figure*}
    \centering
    \plotone{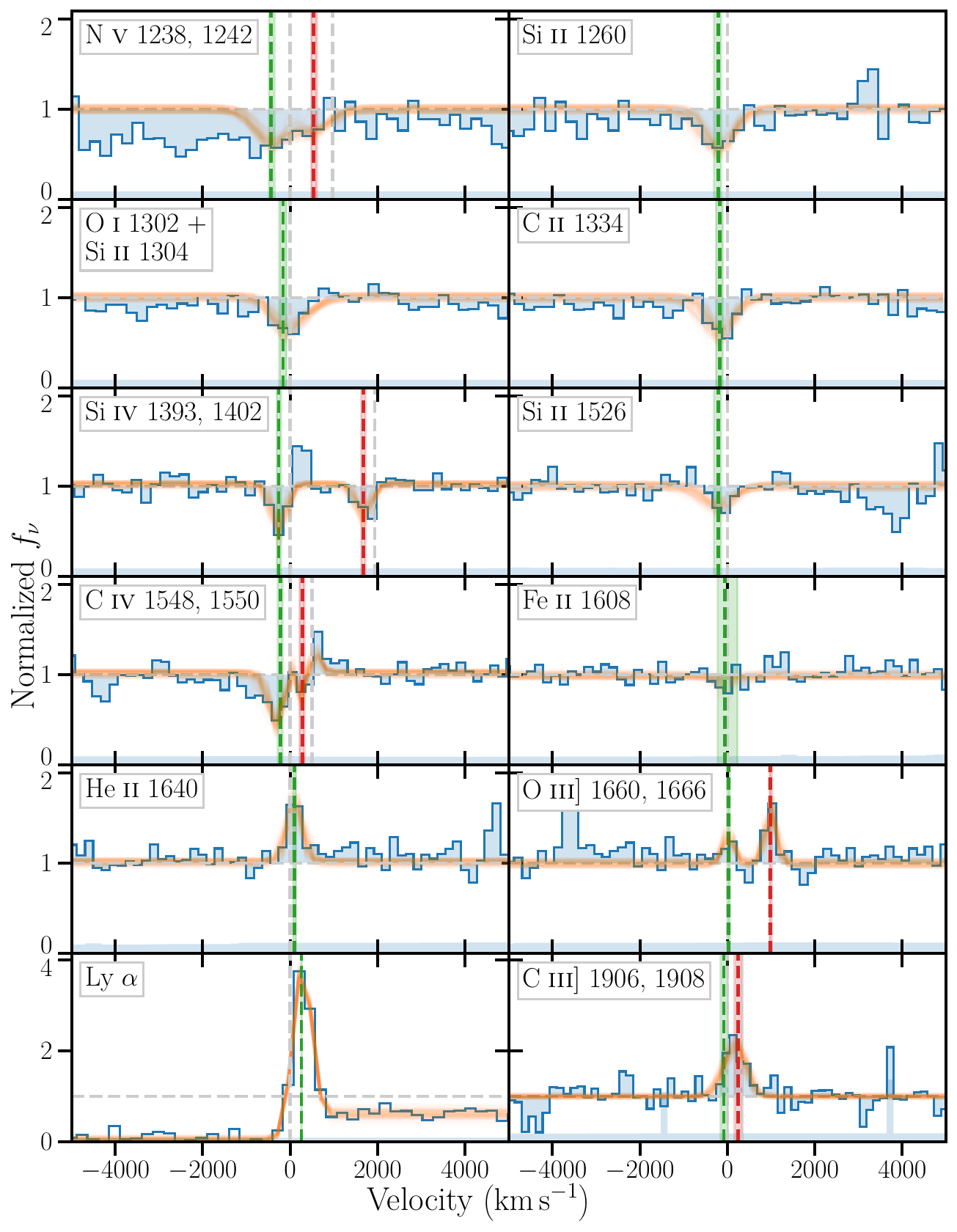}
    \caption{Equivalent plot to Figure \ref{fig:plottransitions}, except for {\tt PypeIt} reductions of the MASQUERADE data.}
    \label{fig:plottransitions2}
\end{figure*}

\begin{figure*}
    \centering
    \plotone{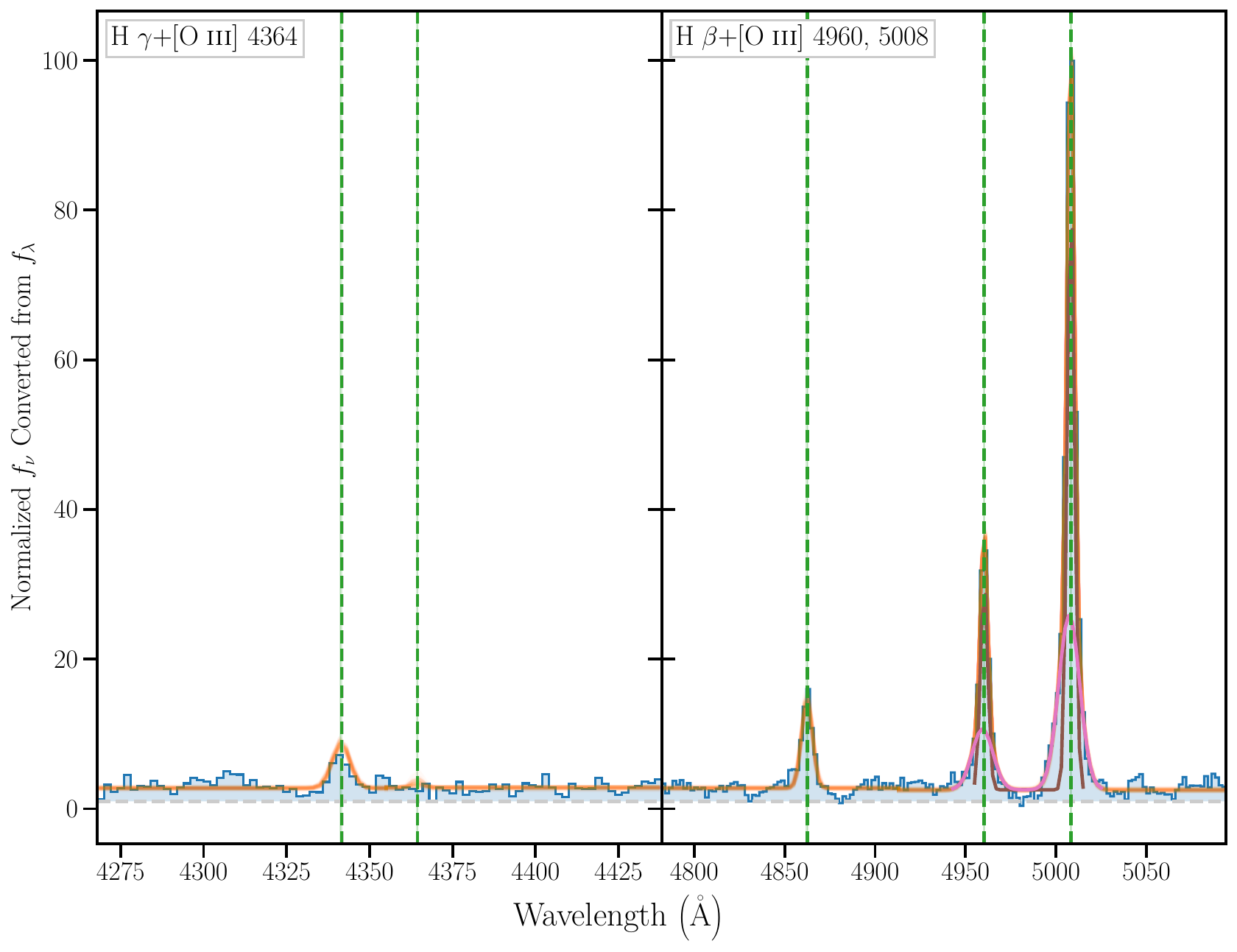}
    \caption{Equivalent plot to Figure \ref{fig:NIRCam_masquerade_stack}, except for weights from {\tt PypeIt}.}
    \label{fig:plotNIRCam2}
\end{figure*}
\FloatBarrier
\section{Table of Galaxies}
\input{appendix}

\end{document}

%% file: macros.tex
\newcommand{\lya}{Ly-$\alpha$~}

\newcommand{\cii}{C~{\sc ii}~}

\newcommand{\civ}{C~{\sc iv}~}

\newcommand{\oi}{O~{\sc i}~}
\newcommand{\oiiisemi}{O~{\sc iii]}~}
\newcommand{\ciiisemi}{C~{\sc iii]}~}
\newcommand{\oiiiforbid}{[O~{\sc iii}]~}

\newcommand{\siii}{Si~{\sc ii}~}

\newcommand{\siiv}{Si~{\sc iv}~}
\newcommand{\heii}{He~{\sc ii}~}
\newcommand{\nv}{N~{\sc v}~}
\newcommand{\feii}{Fe~{\sc ii}~}

\newcommand{\kms}{km s$^{-1}$}

%% file: equivalentwidthdispersion.tex
\begin{deluxetable}{lccccccccc}
\tablecaption{Selected UV Line Transitions and Measured Quantities\label{tab:lines}}
\tablehead{
 & & \multicolumn{4}{c}{MSAExp} & \multicolumn{3}{c}{PypeIt} \\
\colhead{Ion} & \colhead{$\lambda_\mathrm{rest}$} & \colhead{$\sigma_{\rm resel}$\tablenotemark{\small a}}& \colhead{$v$\tablenotemark{\small b}} & \colhead{$\sigma$} & \colhead{$W_r$} & \colhead{$v$\tablenotemark{\small b}} & \colhead{$\sigma$} & \colhead{$W_r$} \\
 & \colhead{\AA} & \colhead{\unit{\km\per\s}} & & \colhead{\unit{\km\per\s}} & \colhead{\AA} & & \colhead{\unit{\km\per\s}} & \colhead{\AA}
}
\decimals
\startdata
\cutinhead{Emission}
Ly\,$\alpha$ & 1215.7  & 196.2 & $246.3_{-7.6}^{+7.9}$ & $200.7_{-6.4}^{+7.1}$ & $11.016_{-1.341}^{+1.711}$ & $258.8_{-8.5}^{+9.5}$ & $203.8_{-7.9}^{+9.0}$ & $12.680_{-1.981}^{+2.866}$ \\
\ion{C}{4}   & 1548.2  & 154.1 & $0.0$\tablenotemark{\small c}                   & $154.1$\tablenotemark{\small c}                 & $1.258_{-0.345}^{+0.354}$ & $0.0$\tablenotemark{\small c}                    & $152.9$\textbf{\tablenotemark{\small c}}                 & $1.197_{-0.362}^{+0.366}$ \\
\ion{C}{4}   & 1550.8  & 153.3 & $0.0$\tablenotemark{\small c}                 & $153.3$\textbf{\tablenotemark{\small c}}                 & $0.929_{-0.265}^{+0.289}$ & $0.0$\tablenotemark{\small c}                   & $152.2$\textbf{\tablenotemark{\small c}}                 & $0.896_{-0.284}^{+0.324}$ \\
\ion{He}{2}  & 1640.5  & 145.0 & $142.6_{-55.3}^{+66.1}$ & $308.0_{-63.3}^{+81.9}$ & $1.811_{-0.399}^{+0.466}$ & $98.7_{-31.3}^{+35.1}$ & $175.7_{-39.2}^{+53.5}$ & $1.619_{-0.318}^{+0.409}$ \\
\ion{O}{3}]  & 1660.8  & 143.2 & $4.2_{-18.7}^{+18.2}$    & $118.6_{-21.4}^{+27.5}$ & $0.530_{-0.104}^{+0.151}$ & $23.4_{-26.0}^{+28.9}$  & $124.7_{-29.5}^{+34.3}$ & $0.579_{-0.149}^{+0.178}$ \\
\ion{O}{3}]  & 1666.2  & 142.7 & $4.2_{-18.7}^{+18.2}$    & $118.6_{-21.4}^{+27.5}$ & $1.058_{-0.183}^{+0.230}$ & $23.4_{-26.0}^{+28.9}$  & $124.7_{-29.5}^{+34.3}$ & $1.087_{-0.269}^{+0.286}$ \\
\ion{C}{3}]  & 1906.7  & 142.4 & $-3.2_{-14.4}^{+12.7}$  & $141.6_{-13.6}^{+23.5}$ & $2.202_{-0.153}^{+0.180}$ & $-80.1_{-75.5}^{+95.8}$ & $201.6_{-41.5}^{+43.0}$ & $1.232_{-0.962}^{+0.877}$ \\
\ion{C}{3}]  & 1908.7  & 142.7 & $-3.2_{-14.4}^{+12.7}$  & $141.6_{-13.6}^{+23.5}$ & $2.371_{-0.158}^{+0.242}$ & $-80.1_{-75.5}^{+95.8}$ & $201.6_{-41.5}^{+43.0}$ & $3.215_{-0.853}^{+1.013}$ \\
\cutinhead{Absorption}
\ion{N}{5}   & 1238.8  & 193.1 & $-445.9_{-38.3}^{+62.3}$ & $345.8_{-55.9}^{+73.6}$ & $1.565_{-0.241}^{+0.249}$ & $-436.2_{-46.1}^{+80.3}$ & $371.6_{-72.3}^{+75.2}$ & $1.499_{-0.272}^{+0.261}$ \\
\ion{N}{5}   & 1242.8  & 192.5 & $-445.9_{-38.3}^{+62.3}$ & $345.8_{-55.9}^{+73.6}$ & $1.011_{-0.166}^{+0.187}$ & $-436.2_{-46.1}^{+80.3}$ & $371.6_{-72.3}^{+75.2}$ & $0.934_{-0.169}^{+0.184}$ \\
\ion{C}{2}   & 1334.5  & 178.9 & $-185.3_{-81.1}^{+63.3}$  & $346.8_{-76.4}^{+89.9}$ & $1.721_{-0.328}^{+0.378}$ & $-170.1_{-94.4}^{+69.4}$ & $298.7_{-64.6}^{+97.5}$ & $1.415_{-0.292}^{+0.341}$ \\
\ion{Si}{2}  & 1260.4  & 189.5 & $-210.6_{-93.3}^{+70.5}$  & $361.6_{-76.9}^{+84.7}$ & $1.488_{-0.312}^{+0.351}$ & $-213.0_{-92.9}^{+67.0}$ & $347.4_{-67.6}^{+83.7}$ & $1.628_{-0.320}^{+0.379}$ \\
\ion{Si}{2}  & 1526.7  & 156.3 & $-210.6_{-93.3}^{+70.5}$  & $361.6_{-76.9}^{+84.7}$ & $1.041_{-0.307}^{+0.346}$ & $-213.0_{-92.9}^{+67.0}$ & $347.4_{-67.6}^{+83.7}$ & $0.998_{-0.310}^{+0.331}$ \\
\ion{Si}{4}  & 1393.8  & 170.6 & $-166.5_{-33.0}^{+18.1}$  & $134.2_{-32.0}^{+38.4}$ & $1.026_{-0.144}^{+0.157}$ & $-261.5_{-46.2}^{+48.7}$ & $169.8_{-31.0}^{+34.1}$ & $1.012_{-0.197}^{+0.195}$ \\
\ion{Si}{4}  & 1402.8  & 169.5 & $-166.5_{-33.0}^{+18.1}$  & $134.2_{-76.9}^{+84.7}$ & $0.733_{-0.130}^{+0.139}$ & $-261.5_{-46.2}^{+48.7}$ & $169.8_{-67.6}^{+83.7}$ & $0.717_{-0.159}^{+0.176}$ \\
\ion{C}{4}   & 1548.2  & 154.2 & $-197.4_{-47.2}^{+39.2}$  & $191.7_{-22.9}^{+28.9}$ & $1.806_{-0.331}^{+0.340}$ & $-222.3_{-63.0}^{+46.8}$ & $204.4_{-26.0}^{+39.6}$ & $1.906_{-0.340}^{+0.346}$ \\
\ion{C}{4}   & 1550.8  & 153.9 & $-197.4_{-47.2}^{+39.2}$  & $191.7_{-22.9}^{+28.9}$ & $1.207_{-0.269}^{+0.303}$ & $-222.3_{-63.0}^{+46.8}$ & $204.4_{-26.0}^{+39.6}$ & $1.390_{-0.311}^{+0.342}$ \\
\ion{O}{1}   & 1302.2  & 183.4 & $-183.7_{-154.7}^{+177.3}$ & $303.0_{-164.6}^{+126.1}$ & $0.609_{-0.292}^{+0.309}$ & $-162.0_{-84.0}^{+81.0}$ & $260.4_{-61.6}^{+84.3}$ & $0.893_{-0.246}^{+0.252}$ \\
\ion{Fe}{2}  & 1608.5  & 147.8 & $165.1_{-171.7}^{+175.3}$ & $338.1_{-128.6}^{+105.2}$ & $0.762_{-0.367}^{+0.362}$ & $-53.9_{-168.5}^{+285.3}$ & $162.8_{-122.0}^{+215.4}$ & $0.320_{-0.219}^{+0.290}$ \\
\enddata
\tablenotetext{a}{\small This column indicates the estimated spectral resolution of the JWST/NIRSpec stack at the fitted wavelength of interest. In cases where the Gaussian line width $\sigma \lesssim \sigma_{\rm resel}$, the line fit is unresolved by the spectrograph.}
\tablenotetext{b}{\small Velocities are relative to [\ion{O}{3}] 4960, 5008.}
\tablenotetext{c}{\small For the emission component of the \ion{C}{4} P-Cygni profile, the velocity centroid is fixed at 0 \kms relative to [\ion{O}{3}] 4960, 5008, and the standard deviation is fixed to the width of one resolution element.}
\label{tab:linemeasurements}
\end{deluxetable}

%% file: optical.tex
\begin{deluxetable*}{cDcCCDC}
    \tablecaption{Selected Optical Line Transitions and Measured Quantities\label{tab:optical}}
    \tablehead{\colhead{Ion}&\twocolhead{$\lambda_\mathrm{rest}$}&\colhead{Component}&\colhead{$v$}&\colhead{$\sigma$}&\twocolhead{$\Delta\lambda$ at $v_{50}$}&\colhead{$W_r$}\\&\twocolhead{\AA}&&\colhead{\unit{\km\per\s}}&\colhead{\unit{\km\per\s}}&\twocolhead{\unit{\km\per\s}}&\colhead{\AA}}
    \decimals
    \startdata
    H\;$\beta$&4862.7 &Single&{-19.8}_{-13.6}^{+13.7}&{188.2}_{-14.5}^{+15.6}&82.5 &{30.1}_{-2.8}^{+3.0}\\
    H\;$\gamma$&4341.7 &Single&{-19.8}_{-13.6}^{+13.7}&{188.2}_{-14.5}^{+15.6}&86.8 &{13.3}_{-1.3}^{+1.4}\\
    {[\ion{O}{3}]}&5008.2 &Narrow&{0.3}_{-2.9}^{+4.4}&{127.6}_{-5.7}^{+7.6}&81.6 &{161.1}_{-15.8}^{+17.1}\\
    {[\ion{O}{3}]}&5008.2 &Wide&{-44.8}_{-18.6}^{+14.5}&{351.0}_{-21.7}^{+24.1}&81.6 &{135.4}_{-15.1}^{+15.0}\\
    {[\ion{O}{3}]}&4960.3 &Narrow&{0.3}_{-2.9}^{+4.4}&{127.6}_{-5.7}^{+7.6}&81.8 &{55.4}_{-5.5}^{+5.9}\\
    {[\ion{O}{3}]}&4960.3 &Wide&{-44.8}_{-18.6}^{+14.5}&{351.0}_{-21.7}^{+24.1}&81.8 &{46.8}_{-5.2}^{+5.1}\\
    {[\ion{O}{3}]}&4364.4 &Single&0.3&127.6&86.5 &{1.7}_{-1.1}^{+1.5}
    \enddata
    \tablecomments{We rounded vacuum wavelengths, velocities, and equivalent widths to one decimal place. For [\ion{O}{3}] 4364, we fixed the velocity and standard deviation to match the narrow components of [\ion{O}{3}] 4960, 5008.}
\end{deluxetable*}

%% file: appendix.tex
\startlongtable
\begin{deluxetable*}{cDDDDDDDDD}
    \tablecaption{Our Galaxies}
    \tablehead{\colhead{Field} & \multicolumn2c{Number} & \multicolumn2c{$z$} & \multicolumn2c{$r_\mathrm{Kron}$} & \multicolumn2c{$M_\mathrm{UV}$} & \multicolumn2c{$\log{M_\star/\mathrm{M_\odot}}$} & \multicolumn2c{$\log{M_\mathrm{h}/\mathrm{M_\odot}}$} & \multicolumn2c{$t_{20}$} & \multicolumn{4}{c}{SNR}\\ & \multicolumn2c{} & \multicolumn2c{} & \multicolumn2c{$\mathrm{mas}$} & \multicolumn2c{mag(AB)} & \multicolumn2c{} & \multicolumn2c{} & \multicolumn2c{Myr} & \multicolumn2c{\texttt{PypeIt}} & \multicolumn2c{\texttt{MSAExp}}}
    \startdata
    \decimals
    J0100+2802 & 30 & 5.9436 & 395 & -20.26 & 8.35 & 11.41 & 33 & 1.68 & 1.98 \\
    J0100+2802 & 452 & 6.6782 & 350 & -18.81 & 7.61 & 11.14 & 8 & 0.25 & 0.55 \\
    J0100+2802 & 812 & 6.7684 & 300 & -20.23 & 9.36 & 11.88 & 139 & 1.45 & 1.54 \\
    J0100+2802 & 1228 & 6.3328 & 328 & -20.73 & 9.45 & 11.92 & 38 & 1.66 & 1.70 \\
    J0100+2802 & 2007 & 6.7696 & 293 & -19.71 & 8.08 & 11.31 & 4 & 0.19 & 0.33 \\
    J0100+2802 & 2084 & 6.7677 & 275 & -19.99 & 8.62 & 11.54 & 36 & 1.27 & 1.46 \\
    J0100+2802 & 2159 & 6.0079 & 336 & -20.20 & 8.41 & 11.44 & 19 & 1.02 & 1.00 \\
    J0100+2802 & 2355 & 6.1824 & 342 & -19.71 & 8.69 & 11.56 & 59 & 0.37 & 0.34 \\
    J0100+2802 & 3783 & 6.1826 & 351 & -20.26 & 9.09 & 11.75 & 19 & 0.67 & 0.69 \\
    J0100+2802 & 4080 & 5.9421 & 360 & -19.64 & 8.99 & 11.70 & 188 & 0.81 & 1.07 \\
    J0100+2802 & 4184 & 6.1866 & 322 & -18.91 & 7.93 & 11.25 & 14 & 0.32 & 0.03 \\
    J0100+2802 & 4222 & 6.3300 & 321 & -21.53 & 9.33 & 11.86 & 37 & 1.56 & 1.66 \\
    J0100+2802 & 5378 & 5.5682 & 555 & -19.29 & 8.54 & 11.49 & 467 & 0.20 & 0.26 \\
    J0100+2802 & 5581 & 5.5095 & 490 & -19.22 & 7.96 & 11.26 & 8 & 0.37 & 0.42 \\
    J0100+2802 & 5595 & 5.9897 & 294 & -19.92 & 8.00 & 11.28 & 15 & 0.60 & 0.33 \\
    J0100+2802 & 5671 & 5.9863 & 355 & -18.92 & 8.64 & 11.54 & 690 & 0.81 & 0.96 \\
    J0100+2802 & 6733 & 6.7681 & 275 & -21.90 & 9.39 & 11.90 & 52 & 5.24 & 5.24 \\
    J0100+2802 & 7194 & 6.1884 & . & . & . & . & . & 0.09 & 0.21 \\
    J0100+2802 & 7230 & 6.8300 & 540 & -19.40 & 8.12 & 11.33 & 34 & 0.06 & 0.23 \\
    J0100+2802 & 7430 & 6.6630 & 328 & -19.18 & 7.97 & 11.27 & 18 & 0.50 & 0.56 \\
    J0100+2802 & 8514 & 6.1819 & 289 & -18.22 & 7.01 & 10.94 & 5 & 0.35 & 0.54 \\
    J0100+2802 & 9273 & 6.1866 & 345 & -20.51 & 8.67 & 11.55 & 19 & 0.34 & 0.41 \\
    J0100+2802 & 9970 & 6.3244 & 477 & -19.39 & 8.44 & 11.46 & 54 & 0.83 & 0.67 \\
    J0100+2802 & 10116 & 5.7808 & 393 & -20.85 & 8.85 & 11.63 & 42 & 1.04 & 1.30 \\
    J0100+2802 & 10235 & 6.3441 & . & . & . & . & . & 0.46 & 0.61 \\
    J0100+2802 & 10287 & 6.7696 & 367 & -19.76 & 7.85 & 11.23 & 26 & 1.23 & 0.48 \\
    J0100+2802 & 10683 & 6.3278 & 329 & -19.06 & 8.01 & 11.28 & 128 & 0.70 & 0.98 \\
    J0100+2802 & 10998 & 6.3259 & 402 & -21.73 & 9.29 & 11.85 & 38 & 1.30 & 1.56 \\
    J0100+2802 & 11362 & 6.3282 & 376 & -19.82 & 7.51 & 11.10 & 4 & 0.45 & 0.54 \\
    J0100+2802 & 11743 & 6.8112 & 289 & -21.27 & 9.61 & 12.01 & 558 & 1.97 & . \tablenotemark{a} \\
    J0100+2802 & 11981 & 6.8218 & . & . & . & . & . & 3.14 & 3.10 \\
    J0100+2802 & 12217 & 6.7731 & . & . & . & . & . & 1.81 & 2.04 \\
    J0100+2802 & 12285 & 6.2028 & 421 & -20.20 & 8.42 & 11.45 & 35 & 1.41 & 1.76 \\
    J0100+2802 & 12643 & 6.5890 & 324 & -18.99 & 8.45 & 11.46 & 555 & 0.32 & 0.30 \\
    J0100+2802 & 12754 & 6.3272 & 370 & -20.14 & 8.80 & 11.61 & 39 & 0.22 & 0.02 \\
    J0100+2802 & 12817 & 5.9065 & 297 & -17.27 & 6.92 & 10.90 & 4 & 0.16 & 0.20 \\
    J0100+2802 & 13590 & 6.1878 & 372 & -18.95 & 8.19 & 11.35 & 39 & 0.47 & 0.47 \\
    J0100+2802 & 13753 & 6.1804 & 357 & -20.19 & 9.02 & 11.71 & 39 & 1.04 & 0.78 \\
    J0100+2802 & 13864 & 6.2763 & 359 & -17.67 & 6.86 & 10.89 & 4 & 0.32 & 0.32 \\
    J0100+2802 & 15162 & 6.1935 & 341 & -19.89 & 8.82 & 11.62 & 41 & 0.43 & 0.54 \\
    J0100+2802 & 15437 & 6.7679 & 555 & -19.45 & 7.84 & 11.22 & 17 & 0.32 & 0.32 \\
    J0100+2802 & 15612 & 6.7639 & . & . & . & . & . & 0.71 & 0.86 \\
    J0100+2802 & 15627 & 6.2040 & 421 & -17.58 & 7.21 & 11.00 & 19 & 0.23 & 0.13 \\
    J0100+2802 & 16184 & 5.7690 & 508 & -18.64 & 7.64 & 11.14 & 33 & 0.24 & 0.33 \\
    J0100+2802 & 16322 & 6.3230 & 323 & -19.45 & 8.42 & 11.45 & 83 & 0.80 & 0.86 \\
    J0100+2802 & 16479 & 6.3338 & 370 & -20.20 & 8.35 & 11.42 & 18 & 1.18 & 1.32 \\
    J0100+2802 & 16759 & 6.3785 & 301 & -20.60 & 8.29 & 11.39 & 36 & 1.39 & 1.17 \\
    J0100+2802 & 17352 & 6.1842 & 508 & -18.50 & 8.28 & 11.39 & 135 & 0.09 & 0.09 \\
    J0100+2802 & 17385 & 6.7707 & . & . & . & . & . & 2.72 & 2.97 \\
    J0100+2802 & 17435 & 6.7532 & 306 & -18.75 & 7.55 & 11.12 & 5 & 0.54 & 0.31 \\
    J0100+2802 & 17492 & 6.7277 & 419 & -19.12 & 7.71 & 11.18 & 24 & 0.64 & 0.67 \\
    J0100+2802 & 17577 & 6.7702 & 334 & -22.09 & 9.86 & 12.12 & 36 & 1.85 & 1.40 \\
    J0100+2802 & 18057 & 6.1932 & 368 & -20.05 & 8.82 & 11.62 & 38 & 0.51 & 0.63 \\
    J0100+2802 & 18061 & 6.3309 & 369 & -20.16 & 8.44 & 11.46 & 37 & 1.39 & 1.54 \\
    J0100+2802 & 18268 & 6.6357 & 348 & -20.23 & 8.24 & 11.38 & 18 & 1.50 & 1.52 \\
    \hline
    J1148+5251 & 2429 & 5.5802 & 332 & -17.66 & 7.99 & 11.27 & 63 & 0.38 & 0.38 \\
    J1148+5251 & 2637 & 6.0372 & 333 & -21.17 & 9.49 & 11.94 & 401 & 1.46 & 1.86 \\
    J1148+5251 & 3052 & 6.4898 & 313 & -18.90 & 7.85 & 11.23 & 536 & 0.37 & 0.59 \\
    J1148+5251 & 3391 & 6.0751 & 370 & -20.07 & 8.83 & 11.62 & 91 & 1.74 & 1.88 \\
    J1148+5251 & 3461 & 6.0386 & 293 & -20.30 & 8.70 & 11.56 & 40 & 1.36 & 1.60 \\
    J1148+5251 & 3607 & 5.4492 & 310 & -18.80 & 8.11 & 11.31 & 153 & 0.59 & 1.07 \\
    J1148+5251 & 3608 & 6.0418 & 403 & -19.15 & 8.29 & 11.39 & 77 & 0.45 & 0.43 \\
    J1148+5251 & 3609 & 6.9081 & 571 & -17.77 & 7.77 & 11.20 & 65 & 0.11 & 0.17 \\
    J1148+5251 & 3781 & 5.9236 & 376 & -20.37 & 9.44 & 11.91 & 112 & 1.00 & 1.90 \\
    J1148+5251 & 3822 & 6.3749 & . & . & . & . & . & 0.11 & 0.24 \\
    J1148+5251 & 4246 & 6.4846 & 427 & -19.69 & 8.47 & 11.47 & 36 & 0.45 & 0.54 \\
    J1148+5251 & 4376 & 6.4881 & 362 & -20.86 & 9.51 & 11.95 & 414 & 0.79 & 0.99 \\
    J1148+5251 & 5123 & 6.4792 & 558 & -18.74 & 8.02 & 11.29 & 35 & 0.34 & 0.41 \\
    J1148+5251 & 6406 & 6.4851 & 407 & -18.40 & 7.38 & 11.06 & 9 & 0.72 & 0.42 \\
    J1148+5251 & 6584 & 6.5805 & 376 & -18.87 & 7.51 & 11.11 & 18 & 0.19 & 0.39 \\
    J1148+5251 & 7217 & 6.4223 & 429 & -18.67 & 7.96 & 11.27 & 56 & 0.31 & 0.50 \\
    J1148+5251 & 8234 & 6.2564 & 287 & -17.09 & 8.51 & 11.48 & 276 & 0.27 & 0.26 \\
    J1148+5251 & 8871 & 6.3881 & 334 & -17.91 & 7.67 & 11.16 & 39 & 0.31 & 0.18 \\
    J1148+5251 & 9232 & 6.0662 & 414 & -16.68 & 6.96 & 10.92 & 7 & 0.19 & 0.23 \\
    J1148+5251 & 9314 & 6.1561 & 328 & -19.82 & 8.82 & 11.62 & 548 & 0.82 & 1.32 \\
    J1148+5251 & 9428 & 6.7687 & 318 & -18.36 & 7.26 & 11.02 & 61 & 0.35 & 0.37 \\
    J1148+5251 & 9799 & 6.4772 & 329 & -18.15 & 8.18 & 11.35 & 79 & 0.04 & 0.19 \\
    J1148+5251 & 10290 & 6.4211 & 389 & -20.13 & 8.77 & 11.60 & 30 & 1.02 & 1.87 \\
    J1148+5251 & 10586 & 5.8172 & 490 & -18.78 & 7.85 & 11.22 & 344 & 0.47 & 0.80 \\
    J1148+5251 & 10727 & 5.7030 & 302 & -18.91 & 7.98 & 11.27 & 42 & 0.92 & 0.53 \\
    J1148+5251 & 11709 & 6.5031 & 303 & -20.81 & 8.94 & 11.68 & 33 & 1.94 & 2.23 \\
    J1148+5251 & 11725 & 6.2062 & 373 & -19.77 & 8.71 & 11.57 & 149 & 0.62 & 0.75 \\
    J1148+5251 & 12115 & 6.9037 & 465 & -19.28 & 8.04 & 11.30 & 36 & 0.88 & 0.79 \\
    J1148+5251 & 12277 & 6.4141 & 283 & -20.88 & 9.63 & 12.01 & 149 & 1.71 & 2.34 \\
    J1148+5251 & 13418 & 5.4283 & 487 & -18.74 & 7.25 & 11.01 & 4 & 0.28 & 0.61 \\
    J1148+5251 & 14101 & 6.8895 & 367 & -19.86 & 8.09 & 11.32 & 17 & 0.97 & 1.08 \\
    J1148+5251 & 14236 & 6.4198 & 406 & -19.29 & 7.47 & 11.09 & 9 & 0.37 & 0.77 \\
    J1148+5251 & 14780 & 6.8873 & 332 & -20.75 & 8.79 & 11.61 & 76 & 1.12 & 1.50 \\
    J1148+5251 & 15411 & 6.9112 & 283 & -21.59 & 9.61 & 12.01 & 65 & 2.42 & 2.42 \\
    J1148+5251 & 16089 & 6.3793 & 284 & -18.73 & 8.44 & 11.46 & 32 & 0.53 & 0.38 \\
    J1148+5251 & 16651 & 6.9033 & 371 & -20.49 & 8.76 & 11.60 & 46 & 0.59 & 0.61 \\
    J1148+5251 & 16736 & 6.9072 & 271 & -20.56 & 8.96 & 11.69 & 68 & 1.40 & 1.31 \\
    J1148+5251 & 16802 & 6.3949 & 339 & -20.87 & 9.01 & 11.71 & 37 & 0.59 & 1.26 \\
    J1148+5251 & 16972 & 6.9049 & 449 & -20.14 & 8.43 & 11.46 & 35 & 1.07 & 1.14 \\
    J1148+5251 & 17076 & 6.4532 & 501 & -19.49 & 7.75 & 11.19 & 17 & 0.08 & 1.41 \\
    J1148+5251 & 17242 & 6.7945 & . & . & . & . & . & 0.71 & 0.82 \\
    J1148+5251 & 17433 & 6.2556 & 521 & -16.41 & 7.65 & 11.15 & 49 & 0.16 & 0.15 \\
    J1148+5251 & 17783 & 5.9258 & 417 & -17.59 & 7.17 & 10.98 & 38 & 0.24 & 0.30 \\
    J1148+5251 & 18178 & 5.7293 & 367 & -17.23 & 7.06 & 10.95 & 16 & 1.35 & 1.31 \\
    J1148+5251 & 18852 & 5.4387 & 382 & -16.88 & 6.92 & 10.90 & 5 & 0.13 & 0.09 \\
    J1148+5251 & 18887 & 6.0107 & 339 & -18.79 & 8.37 & 11.42 & 695 & 0.08 & 0.09 \\
    J1148+5251 & 19229 & 6.7494 & 412 & -19.10 & 7.75 & 11.19 & 17 & 0.49 & 0.65 \\
    J1148+5251 & 19514 & 5.6863 & 303 & -18.06 & 11.06 & 12.53 & 701 & 0.40 & 0.81 \\
    J1148+5251 & 19692 & 6.9016 & 336 & -20.18 & 8.62 & 11.54 & 36 & 1.54 & 1.49
    \enddata
    \tablenotetext{a}{\texttt{MSAExp} did not extract this galaxy because it shares a slit with the brighter galaxy 11981.}
\end{deluxetable*}